\documentclass[
 preprint,
superscriptaddress,
nofootinbib,
 amsmath,amssymb,
 aps,
prd
]{revtex4-2}
\pdfoutput=1

\usepackage{amsbsy}

\usepackage{graphicx}
\usepackage{dcolumn}
\usepackage{bm}
\usepackage{setspace}
\usepackage{hyperref}
\hypersetup{
colorlinks=true,
linkcolor=blue,
citecolor=blue,
linktocpage
}

\usepackage{color}
\usepackage{braket}
\usepackage[normalem]{ulem}

\graphicspath{{./}{./figs/}}

\newcommand{\dd}{\mathrm{d}}

\newcommand{\del}{\partial}
\newcommand{\p}{\partial}
\newcommand{\rmmax}{\mathrm{max}}
\newcommand{\rmmin}{\mathrm{min}}

\newcommand{\frakI}{{\mathfrak{i}}}
\newcommand{\frakJ}{{\mathfrak{j}}}

\newcommand{\onmom}[2]{\ve{p}_{#1}^{#2}}

\newcommand{\rmRe}{\mathrm{Re}}
\newcommand{\rmIm}{\mathrm{Im}}
\newcommand{\PV}{\mathrm{P.V.}}

\newcommand{\ve}[1]{\textbf{#1}}
\newcommand{\Dmax}{\Delta_{\mathrm{max}}}

\newcommand{\imax}{\mathfrak{i}_{\max}}
\newcommand{\LambdaUV}{\Lambda_{\textrm{UV}}}
\newcommand{\be}{\begin{equation}}
\newcommand{\ee}{\end{equation}}
\newcommand{\ba}{\begin{aligned}}
\newcommand{\ea}{\end{aligned}}

\newcommand{\fr}{\frac}

\newcommand{\half}{\frac{1}{2}}

\newcommand{\Lcal}{{\mathcal L}}
\newcommand{\Mcal}{{\mathcal M}}

\newcommand{\Ocal}{{\mathcal O}}

\newcommand{\ifrak}{{\mathfrak{i}}}
\newcommand{\jfrak}{{\mathfrak{j}}}
\newcommand{\de}{\delta}
\newcommand{\De}{\Delta}

\newcommand{\<}{\langle}
\renewcommand{\>}{\rangle}
\newcommand{\ra}{\rightarrow}

\newcommand{\lra}{\leftrightarrow}

\newcommand{\order}{O}

\makeatletter
\def\blfootnote{\gdef\@thefnmark{}\@footnotetext}
\makeatother

\makeatletter
\def\l@subsubsection#1#2{}
\makeatother

\begin{document}
\count\footins = 1000

\raggedbottom

\title{
Towards a nonperturbative construction of the \texorpdfstring{$S$}{S}-matrix}

\author{Brian Henning}
\affiliation{Institute of Physics, \'Ecole Polytechnique F\'ed\'erale de Lausanne (EPFL), CH-1015 Lausanne, Switzerland}
\affiliation{D\'{e}partment de Physique Th\'{e}orique, Universit\'{e} de Gen\`{e}ve, CH-1211 Gen\`{e}ve, Switzerland}

\author{Hitoshi Murayama}
\affiliation{Department of Physics, University of California, Berkeley, CA 94720, USA}
\affiliation{Kavli Institute for the Physics and Mathematics of the Universe (WPI), University of Tokyo, Kashiwa 277-8583, Japan}
\affiliation{Ernest Orlando Lawrence Berkeley National Laboratory, Berkeley, CA 94720, USA}

\author{Francesco Riva}
\affiliation{D\'{e}partment de Physique Th\'{e}orique, Universit\'{e} de Gen\`{e}ve, CH-1211 Gen\`{e}ve, Switzerland}

\author{Jedidiah O. Thompson}
\affiliation{Stanford Institute for Theoretical Phyiscs, Stanford University, Stanford, CA 94305, USA}

\author{Matthew T.\ Walters}
\affiliation{Institute of Physics, \'Ecole Polytechnique F\'ed\'erale de Lausanne (EPFL), CH-1015 Lausanne, Switzerland}
\affiliation{D\'{e}partment de Physique Th\'{e}orique, Universit\'{e} de Gen\`{e}ve, CH-1211 Gen\`{e}ve, Switzerland}

\begin{abstract}
We present a nonperturbative recipe for directly computing the $S$-matrix in strongly-coupled QFTs. The method makes use of spectral data obtained in a Hamiltonian framework and can be applied to a wide range of theories, including potentially QCD. We demonstrate the utility of this prescription in the specific example of the 2+1d $O(N)$ model at large $N$, using
energy eigenstates computed with Hamiltonian truncation to reproduce the full $2 \to 2$ scattering amplitude for arbitrary (complex) center-of-mass energy.
\end{abstract}

\maketitle
\pagebreak
\tableofcontents

\section{Introduction} \label{sec:intro}

One of the most useful observables in quantum field theory (QFT) is the $S$-matrix: the probability amplitude for a scattering process to occur,
\begin{equation}
    S_{\beta \alpha} \equiv \langle \beta; \mathrm{out} | \alpha; \mathrm{in} \rangle \equiv \langle \psi_\beta^- | \psi_\alpha^+ \rangle,
    \label{eq:SmatDef}
\end{equation}
where the labels ``in'' and ``out'' refer to asymptotic states that are respectively defined in terms of their properties at negative and positive temporal infinity, and $\alpha, \beta$ are multi-labels for the quantum numbers and momenta of all incoming and outgoing particles.

The most well-known technique to calculate $S$-matrix elements is via Feynman diagrams in the context of perturbative QFT. While such an approach has clearly been very successful in many applications, it is fundamentally limited to weakly-interacting systems. In this paper, we wish to focus on a broader question: How can we compute the $S$-matrix in a strongly-coupled theory for which no perturbative expansion exists?

While there are many tools for studying strongly-coupled physics, we currently have no nonperturbative method to directly compute scattering amplitudes. Arguably the most successful tool is lattice Monte Carlo, which is formulated in Euclidean spacetime and therefore does not have direct access to dynamical processes such as scattering. Lattice calculations have been used to indirectly extract partial scattering information from observables such as the finite-volume mass spectrum and operator matrix elements~\cite{Luscher:1986pf,Luscher:1990ux,Luscher:1991cf,Lellouch:2000pv,Aoki:2012tk} (see~\cite{Briceno:2017max} for a recent review), though such methods are limited to the elastic regime (i.e.~energies below multi-particle thresholds). There has been exciting recent progress towards extracting scattering amplitudes from Euclidean correlation functions computed on the lattice~\cite{Hansen:2017mnd,Bulava:2019kbi}, although this approach faces a numerically ill-posed inverse problem in the analytic continuation to Lorentzian signature. Furthermore, there are many QFTs for which no lattice formulation is known, most notably chiral gauge theories such as the Standard Model, and there is currently no existing lattice approach to dynamics at finite temperature or chemical potential, which have direct relevance to heavy-ion experiments and studies of quark-gluon plasma.

In contrast, Hamiltonian-based techniques are formulated in Minkowski spacetime and have direct access to real-time dynamics. For instance, in the specific technique of Hamiltonian truncation -- which we consider in this work -- one reduces the continuum theory's Hamiltonian to a finite-dimensional (matrix) approximation, then diagonalizes this matrix numerically to obtain a discrete, finite set of approximate energy eigenstates (see~\cite{James:2017cpc,Fitzpatrick:2022dwq} for recent overviews). These states have been used to study many Lorentzian observables (such as spectral densities~\cite{Katz:2016hxp}, form factors~\cite{Chen:2021bmm}, and finite-temperature dynamics~\cite{Kukuljan:2018whw,Delacretaz:2022ojg}), but do not have an immediately clear connection to asymptotic states.

This discrete spectrum is a common denominator of numerical techniques and formally prevents the identification of distinct in- and out-states, thus representing the main obstacle to computing the $S$-matrix.
Indeed, most numerical methods are formulated in finite volume, where the finite-size box in position space literally removes the notion of asymptotic infinity. Sometimes the ``box'' resulting from discretization isn't as physically transparent, as is the case for the specific Hamiltonian truncation method employed in this work, which is technically formulated at infinite volume. Nevertheless, the inescapable point is that any finite approximation to the continuum QFT will develop effective UV and IR scales governing the resolving power and preventing a simple identification of asymptotic states.\footnote{
This can be seen more concretely via the Lippmann-Schwinger definition of asymptotic states,
\begin{equation}\label{eq:LS3}
  \ket{\psi_{\alpha}^{\pm}} = \ket{\phi_{\alpha}} + \frac{1}{E_{\alpha} - H_0 \pm i\epsilon} V \ket{\psi_{\alpha}^{\pm}},
\end{equation}
where $|\phi_\alpha\>$ is a free state with the same energy $E_\alpha$ as the interacting eigenstates $|\psi_\alpha^\pm\>$; such a state is guaranteed to exist in the continuum. The $\pm i\epsilon$ fixes the boundary condition $\ket{\psi_{\alpha}^{\pm}}\to \ket{\phi_{\alpha}}$ as $t\to\mp\infty$ (for a nice discussion, see e.g.\ Ref.~\cite{FroissartOmnes}).
In the discretized case, the eigenvalues of $H_0$ generically differ from those of $H = H_0+V$, eliminating the $\pm i \epsilon$ and thus the notion of distinct in- and out-states.}

Our strategy to circumvent this issue is to use the Lehmann-Symanzik-Zimmermann (LSZ) reduction formula, which relates asymptotic states to interpolating fields. For example, the \(2\to 2\) $S$-matrix element of spinless particles \(\braket{\mathbf{p}_3,\mathbf{p}_4;\text{out}|\mathbf{p}_1,\mathbf{p}_2;\text{in}}\) is given by the Fourier transform of the time-ordered correlator
\begin{equation}\label{eq:fundamental_correlator}
  (\Box_3 + m^2)(\Box_2 + m^2)\<\mathbf{p}_4|T\{\phi(x_3)\phi(x_2)\}|\mathbf{p}_1\>.
\end{equation}
To evaluate this correlation function, we make use of the fact that the finite set of approximate eigenstates obtained with Hamiltonian truncation, \(H |\psi_\alpha\> = E_\alpha |\psi_\alpha\>\), provides an approximate resolution of the identity:
\begin{equation}\label{eq:approx_identity}
  1 \approx \sum_{\alpha=1}^{\alpha_{\text{max}}} |\psi_\alpha\> \<\psi_\alpha|,
\end{equation}
which we insert between \(\phi(x_3)\) and \(\phi(x_2)\) for the two different time-orderings in Eq.~\eqref{eq:fundamental_correlator} in order to obtain an $S$-matrix element.  Heuristically, this approach allows us to connect asymptotic states defined in a continuum at infinity (the external particle energies can take continuum values), with discrete states computed in the finite ``box.''

\begin{figure}[t]
    \centering
    \includegraphics[width=\linewidth]{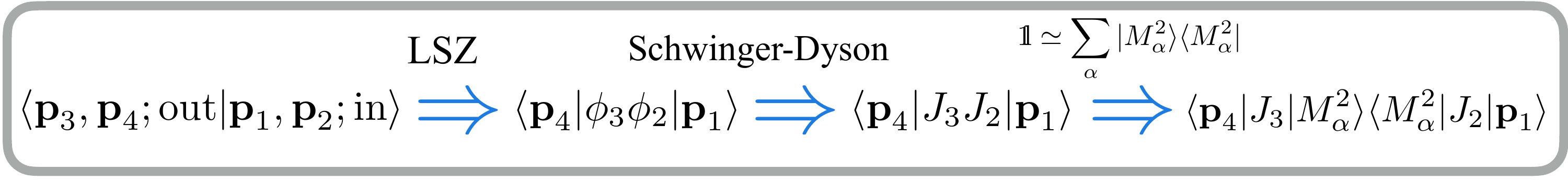}
    \caption{Schematic representation of the steps taken in this work to translate a scattering amplitude into a smooth object computable with knowledge of the discrete spectrum; here $\phi_2=\phi(x_2)$, etc.} \label{fig:Flow}
\end{figure}

However, there is a crucial difficulty in implementing this procedure: the exact zeroes associated with the $(\square + m^2)$ factors in \eqref{eq:fundamental_correlator} are supposed to cancel with exact poles in the correlation function. Because we are computing the correlator with an approximate, discrete spectrum, in practice we only reproduce approximate poles which are unable to cancel the exact zeroes, and the resulting expression naively vanishes for on-shell external particles. We circumvent this issue by using Schwinger-Dyson equations to rewrite \eqref{eq:fundamental_correlator} as a correlator involving sources (such as $J \sim \lambda \phi^3$ for a $\lambda \phi^4$ interaction) and contact terms. The result is a smooth expression for the scattering amplitude, written entirely in terms of matrix elements of $J$ between numerically computed eigenstates of the full Hamiltonian.\footnote{The Euclidean version of this correlation function was also proposed in~\cite{Bulava:2019kbi} as a means of studying scattering amplitudes with lattice field theory.}
This approach, outlined schematically in Fig.~\ref{fig:Flow}, provides a general recipe for directly computing nonperturbative scattering amplitudes in both the elastic and inelastic regions.

\begin{figure}[t]
    \centering
    \includegraphics[width=0.6\linewidth]{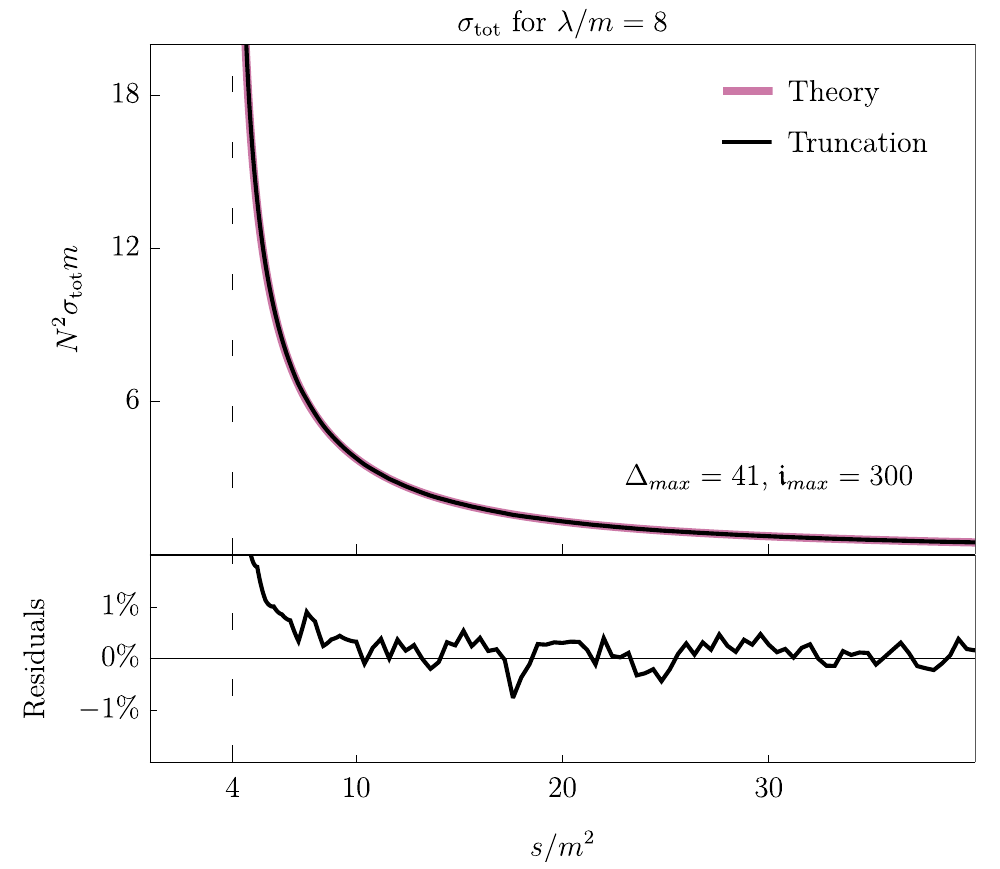}
    \caption{The total cross section for $\phi_i \phi_i \to \phi_i \phi_i$ scattering at strong coupling in the $O(N)$ model at large $N$ in 2+1d.  The pink curve is the exact theoretical answer (obtained from Eq.~\eqref{eq:amp_ON_all_flavors} with $i = j = k = l$) and the black line is a numerical computation performed using the techniques discussed in this paper.  The two curves are in very good agreement, indicating that this method for computing $S$-matrix elements is capable of probing strong coupling. For a much more detailed discussion of these results see Sec.~\ref{sec:ONmodel}.}
    \label{fig:crossSection}
\end{figure}

As a proof of concept, we successfully implement this procedure for a strongly-coupled QFT: the 2+1d $O(N)$ model at large $N$. Using approximate eigenstates obtained with a particular implementation of Hamiltonian truncation known as lightcone conformal truncation~\cite{Katz:2016hxp} (see~\cite{Anand:2020gnn} for a pedagogical introduction), we compute the nonperturbative $2 \to 2$ scattering amplitude over a wide range of energies. Because this theory is exactly solvable in the large-$N$ limit, we can compare our numerical results to known theoretical ones, as shown in Fig.~\ref{fig:crossSection} for the resulting total cross section. With minimal computational cost ($\sim20$ seconds on a typical laptop), we reproduce the exact expression to $\lesssim 1\%$ precision. It is important to note that the exact solvability of this model does not play a significant role in the truncation calculation itself, and this technique can be implemented in more general strongly-interacting QFTs.

There has been other recent work also focused on connecting Hamiltonian truncation methods with the $S$-matrix. In~\cite{Gabai:2019ryw}, the finite-volume spectrum of two-particle states computed with truncation was used to extract the scattering phase in the elastic regime, for the specific case of 1+1d Ising field theory. These elastic results were then combined with unitarity and analyticity assumptions to determine the analytic continuation of the $2\ra2$ $S$-matrix element for arbitrary complex $s$. In~\cite{Chen:2021pgx}, Hamiltonian truncation was used to compute two-particle form factors in 1+1d $\phi^4$ theory, which were combined with $S$-matrix bootstrap techniques developed in~\cite{Karateev:2019ymz} to obtain very restrictive bounds on the scattering amplitude in the elastic regime. In addition to truncation methods, other approaches have been developed in recent years to study dynamical processes nonperturbatively, including the $S$-matrix bootstrap~\cite{Kruczenski:2022lot}, tensor networks~\cite{Banuls:2019rao,Rigobello:2021fxw}, and quantum simulation~\cite{Bauer:2022hpo}.

The remainder of this paper is divided as follows.  In Sec.~\ref{sec:exec_summ} we   discuss the difficulties associated with computing amplitudes using a Hamiltonian approximation scheme, and we show how to circumvent them.
As a demonstration, in Sec.~\ref{sec:ONmodel} we apply this to the $O(N)$ model. We conclude in Sec.~\ref{sec:discussion} with a discussion of our results and their broader applications.
In the interest of efficient presentation, we have split some useful results off into appendices.  In App.~\ref{app:ON_truncation_detail} we collect technical details of the numerical truncation calculation done in this paper, and in App.~\ref{app:LSZ_alt_derivation} we give an alternate derivation of the key formula of Sec.~\ref{sec:exec_summ}.

\section{\texorpdfstring{$S$}{S}-matrix elements from energy eigenstates} \label{sec:exec_summ}
\label{sec:Smat_from_estates}

For concreteness we consider $2 \to 2$ scattering of a single spin-0 particle associated with the field $\phi$. Generalizations to distinct particles and particles with spin are straightforward.\footnote{The case where the asymptotic particle is a bound state that is \textit{not} interpolated by the fundamental field $\phi$ requires a different treatment from the method outlined in this section, particularly in regard to the use of the equations of motion. We comment on this further in Sec.~\ref{sec:discussion}.}
As discussed in the introduction, our starting point is to use LSZ to reduce only two of the external particles, which results in the expression:\footnote{In this work $\langle \ve{p}_3, \ve{p}_4; \mathrm{out} | \ve{p}_1, \ve{p}_2; \mathrm{in} \rangle$ refers solely to the connected part of the $S$-matrix.}
\begin{equation} \label{eq:LSZ2part}
\ba
    \langle \ve{p}_3, \ve{p}_4; \mathrm{out} | \ve{p}_1, \ve{p}_2; \mathrm{in} \rangle &= \bigg( \frac{i}{\sqrt{Z}} \bigg)^2 \! \int \! \dd x_2 \, \dd x_3 \, e^{i (p_3 \cdot x_3 - p_2 \cdot x_2)} \\
    &\qquad \qquad \times (\square_3 + m^2) (\square_2 + m^2) \langle \ve{p}_4 | T \{ \phi(x_3) \phi(x_2) \} | \ve{p}_1 \rangle.
\ea
\end{equation}
The advantage of keeping one external particle in the initial and final states---as opposed to reducing all four particles---is twofold.  First, stable one-particle states are trivially asymptotic:  $|\ve{p}; \mathrm{in} \rangle = |\ve{p}; \mathrm{out} \rangle = |\ve{p} \rangle$. Moreover, on a practical level, the single-particle states are easy to identify in the truncated spectrum of $H$ because they are isolated from the multi-particle continuum.

The correlator on the RHS of Eq.~\eqref{eq:LSZ2part} can be evaluated nonperturbatively by inserting a complete set of energy eigenstates. This is also true if we insert a discrete set of \textit{approximate} energy eigenstates, so long as those eigenstates are computed nonperturbatively, e.g.\ with a method such as Hamiltonian truncation. Naturally, this will only be able to reproduce at best \textit{approximate} poles around $p_2^2 = p_3^2 = m^2$.  According to Eq.~\eqref{eq:LSZ2part}, however, we must then multiply by \textit{exact} zeroes $(p_2^2 - m^2)(p_3^2 - m^2)$ in order to obtain the scattering amplitude.  This type of delicate cancellation is numerically unstable, and is one reason why such a technique has not been implemented in the past.

Fortunately, we can avoid this issue and enforce the cancellation exactly, even with an approximate spectrum, by using the equations of motion at the operator level,
\begin{equation} \label{eq:eom_phi}
    \frac{\delta S}{\delta \phi(x)} = -(\square + m^2) \phi(x) + J(x),
\end{equation}
where $J$ is the source for $\phi$ (e.g.\ $J = - \frac{\lambda}{3!} \phi^3$ for $V(\phi) = \frac{\lambda}{4!} \phi^4$). We can then relate correlation functions involving $(\Box + m^2) \phi$ to those involving $J$ via Schwinger-Dyson equations,\footnote{The Schwinger-Dyson equations stem from invariance of the path integral under variations in $\phi$,
\begin{equation}
\int_{\phi_{\textrm{in}}}^{\phi_{\textrm{out}}} \!\!\! \mathcal{D}\phi \, \fr{\de}{\de\phi} \bigg( e^{iS} \Ocal_1(x_1) \cdots \Ocal_n(x_n) \bigg) = 0, \nonumber
\end{equation}
resulting in the general relation
\begin{equation}
\<\psi_{\textrm{out}}|T \{\fr{\de S}{\de\phi(x)} \Ocal_1(x_1) \cdots \Ocal_n(x_n)\}|\psi_{\textrm{in}}\> = i \sum_i \<\psi_{\textrm{out}}|T \{\Ocal_1(x_1) \cdots \fr{\de \Ocal_i}{\de\phi(x)} \cdots \Ocal_n(x_n)\}|\psi_{\textrm{in}}\>. \nonumber
\end{equation}
Note that the \(\Box\) inside \(\frac{\delta S}{\delta\phi}\) is understood to act from outside the time-ordered product---the basic reason being that \(\Box\) can be brought outside the path integral, and the path integral computes time-ordered correlation functions.
}
\begin{equation}
    \langle \ve{p}_4 | T \{ \frac{\delta S}{\delta \phi(x_3)} \frac{\delta S}{\delta \phi(x_2)} \} | \ve{p}_1 \rangle = i \, \< \ve{p}_4 | \frac{\delta^2 S}{\delta \phi(x_3) \delta \phi(x_2)} | \ve{p}_1 \>.
    \label{eq:DSeqn}
\end{equation}
The left-hand side of~\eqref{eq:DSeqn} contains the correlator appearing in the partially-reduced LSZ formula~\eqref{eq:LSZ2part}, and all remaining terms can be rewritten in terms of the source $J$, resulting in the relation\footnote{For clarity, we have suppressed an additional disconnected term $-i(\Box_2 + m^2)\de^d(x_3 - x_2)\<\ve{p}_4|\ve{p}_1\>$ on the right-hand side which does not contribute to the resulting $S$-matrix element. }
\begin{equation} \label{eq:SDforLSZcorrelator}
\ba
    &(\Box_3 + m^2) (\Box_2 + m^2) \langle \ve{p}_4 | T \{\phi(x_3) \phi(x_2) \} | \ve{p}_1 \rangle \\
    & \hspace{2.5cm} = \langle \ve{p}_4 | T \{J(x_3) J(x_2)\} | \ve{p}_1 \rangle - i \delta^d (x_3 - x_2) \langle \ve{p}_4 | J'(x_2) | \ve{p}_1 \rangle,
\ea
\end{equation}
where the contact term $J' \equiv \frac{\del J}{\del \phi}$ (e.g.\ $J' = - \frac{\lambda}{2} \phi^2$ for $V(\phi) = \frac{\lambda}{4!} \phi^4$) arises from functional differentiation \(\frac{\delta J(x_2)}{\delta \phi(x_3)} \equiv J'(x_2) \delta^d(x_3-x_2)\). The crucial advantage of this relation is that the quantities appearing on the right-hand side are manifestly smooth when particles 2 and 3 are on-shell, contrary to Eq.~\eqref{eq:LSZ2part}. As a result, Eq.~\eqref{eq:SDforLSZcorrelator} provides a numerically stable version of the LSZ reduction formula,
\begin{equation} \label{eq:LSZinTermsOfJJ}
{
\ba
&\<\ve{p}_3,\ve{p}_4;\textrm{out}|\ve{p}_1,\ve{p}_2;\textrm{in}\> =\\
& \,  - \fr{1}{Z} \bigg[ \int \! \dd x_2 \, \dd x_3 \, e^{i(p_3\cdot x_3 - p_2\cdot x_2)} \<\ve{p}_4|T \{J(x_3) J(x_2) \}|\ve{p}_1\> - i\!\int \! \dd x_2 \, e^{i(p_3-p_2)\cdot x_2} \<\ve{p}_4|J'(x_2)|\ve{p}_1\> \bigg].
\ea
}
\end{equation}

\begin{figure}
    \centering
    \includegraphics[width=0.9\textwidth]{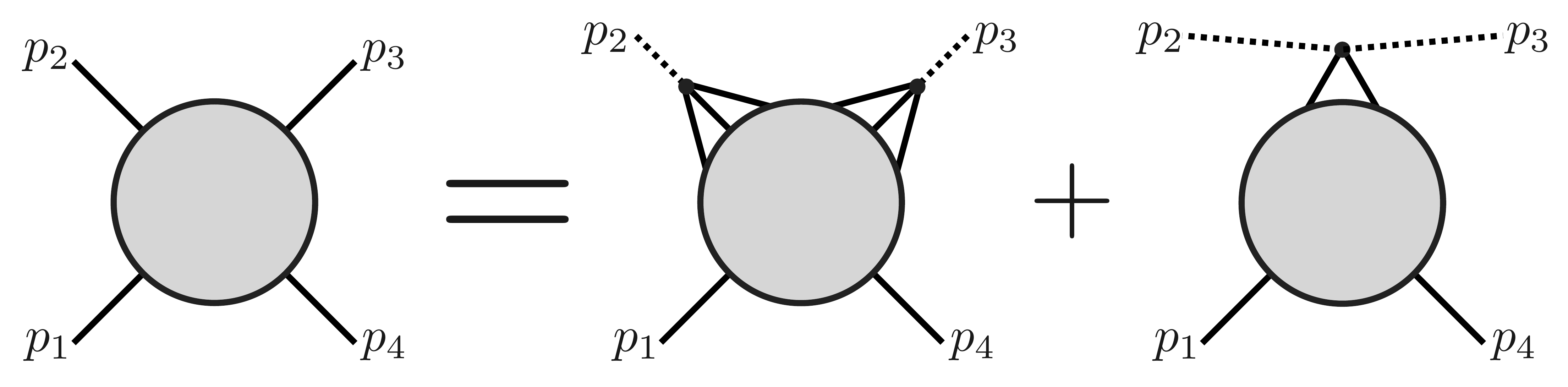}
    \caption{\footnotesize{Schematic representation of Eq.~\eqref{eq:LSZinTermsOfJJ} for the case of $\phi^4$ theory. All contributions to the amplitude (left diagram) either involve particles 2 and 3 contracting with different interaction vertices (middle diagram) or the same vertex (right diagram). The first case results in two insertions of $J = -\frac{\lambda}{3!}\phi^3$, and the second case results in a single insertion of $J' = -\frac{\lambda}{2}\phi^2$. The external propagators for particles 2 and 3 (dashed lines) exactly cancel the explicit factors of $(\Box+m^2)$ in the original LSZ formula~\eqref{eq:LSZ2part}.}}
    \label{fig:SchwingerDysonDiagram}
\end{figure}

Eq.~\eqref{eq:LSZinTermsOfJJ} is nonperturbative and satisfied in general theories, but we can gain some intuition for its structure with the simple diagrammatic interpretation shown in Fig.~\ref{fig:SchwingerDysonDiagram} for the case of a $\phi^4$ interaction.\footnote{In theories with a nonzero correction $\de m^2$ to the bare mass, note that $J$ contains an additional term $J \supset \de m^2 \phi$, which we have suppressed in Fig.~\ref{fig:SchwingerDysonDiagram} for simplicity.}
In any contribution to the $S$-matrix element, particles 2 and 3 can either interact at different vertices or the same vertex, corresponding respectively to the $JJ$ and $J'$ terms. This structure is easily seen at all orders in perturbation theory and holds nonperturbatively due to the Schwinger-Dyson equations (see App.~\ref{app:LSZ_alt_derivation} for an alternative derivation in a canonical formulation).

Given a set of approximate energy eigenstates $| M_\alpha^2 ; \ve{p} \>$, labeled by their total spatial momentum $\ve{p}$ and invariant mass $M_\alpha^2$, we can evaluate Eq.~\eqref{eq:LSZinTermsOfJJ} by inserting them as intermediate states. For example, in the case $t_2 < t_3$ we have
\begin{equation}
    \langle \ve{p}_4 | J(x_3) J(x_2) | \ve{p}_1 \rangle = \sum_\alpha \int_{\ve{p}} \langle \ve{p}_4 | J(x_3) | M_\alpha^2 ; \ve{p} \rangle \langle M_\alpha^2 ; \ve{p} | J(x_2) | \ve{p}_1 \rangle,
\end{equation}
with a similar expression for $t_2 > t_3$.
Because we have inserted eigenstates of the momentum operator $P^\mu$, we can translate $J(x) = e^{i P \cdot x} J(0) e^{-i P \cdot x}$ to obtain
\begin{equation}
    \langle \ve{p}_4 | J(x_3) J(x_2) | \ve{p}_1 \rangle = \sum_\alpha \int_{\ve{p}} e^{i (p_4 - p) \cdot x_3 + i (p - p_1) \cdot x_2} \langle \ve{p}_4 | J(0) | M_\alpha^2 ; \ve{p} \rangle \langle M_\alpha^2 ; \ve{p} | J(0) | \ve{p}_1 \rangle.
\end{equation}
Substituting this expression into~\eqref{eq:LSZinTermsOfJJ} and integrating over $x_2$ and $x_3$ we find
\begin{equation} \label{eq:GeneralAmp}
\boxed{
\ba
\Mcal(s,t) &= \fr{1}{Z} \Bigg[ \sum_\alpha \bigg( \<\ve{p}_4|J(0)|M_\alpha^2 ; \ve{p}_1 + \ve{p}_2 \>\<M_\alpha^2 ; \ve{p}_1 + \ve{p}_2 | J(0) |\ve{p}_1\> \fr{1}{M_\alpha^2-s-i\epsilon} \\
& \, + \<\ve{p}_4|J(0)|M_\alpha^2 ; \ve{p}_1 - \ve{p}_3 \>\<M_\alpha^2 ; \ve{p}_1 - \ve{p}_3 | J(0) |\ve{p}_1\> \fr{1}{M_\alpha^2-t-i\epsilon} \bigg) + \<\ve{p}_4|J'(0)|\ve{p}_1\> \Bigg],
\ea
}
\end{equation}
where the scattering amplitude $\Mcal$ is simply the $S$-matrix element with the momentum-conserving delta function factored out,
\begin{equation}
    \<\ve{p}_3,\ve{p}_4;\textrm{out}|\ve{p}_1,\ve{p}_2;\textrm{in}\> \equiv (2 \pi)^d \delta^d(p_4 + p_3 - p_2 - p_1) \, i \Mcal(s, t),
\end{equation}
and
$s \equiv (p_1 + p_2)^2$ and $t \equiv (p_1 - p_3)^2$ are the Mandelstam variables. The $s$- and $t$-terms in~\eqref{eq:GeneralAmp} (crossing symmetric under $s \leftrightarrow t$, or $p_2 \leftrightarrow - p_3$)  come from the two different choices of time-ordering ($t_2 < t_3$ and $t_2 > t_3$, respectively) in~\eqref{eq:LSZinTermsOfJJ}. The remaining crossing symmetry ($s,t \lra u$) is hidden but still present in this expression, and can be made manifest by choosing to reduce out a different pair of particles in the LSZ reduction formula (for further discussion on this point see Sec.~\ref{sec:discussion}).

Eq.~\eqref{eq:GeneralAmp} encodes the fundamental---and conceptually simple---idea of this work: LSZ relates $S$-matrix elements to correlation functions, which in turn can be evaluated by inserting the (approximate) energy eigenstates $| M_\alpha^2 ; \ve{p} \>$. Importantly, \eqref{eq:GeneralAmp} can be fully evaluated using only information from these states. The external states $|\ve{p}_{1,4}\>$ are simply the one-particle energy eigenstate evaluated at two different spatial momenta and can be easily identified in the spectrum of eigenstates. In addition, the field strength renormalization $Z$ can be computed directly from the overlap of this one-particle state with the field $\phi$,
\be
Z \equiv \big| \<\phi(0)|\ve{p}\> \big|^2.
\ee

Notice that, for a real source, the last term $ \langle \ve{p}_4 | J'(0) | \ve{p}_1 \rangle$ is purely real. The imaginary part of the amplitude thus arises purely from the $J J$ terms, which explicitly come from propagating, on-shell intermediate states. Additionally---again, for a real source---the matrix elements appearing in the $JJ$ term are purely real. Then for fixed\footnote{\label{ftntan}Here $u$ is required to take physical values since particles $1$ and $4$ are physical external states; $s$ and $t$ can be simultaneously complex, so long as $u$ is real and negative. In the complex energy plane, the amplitude will exhibit non-analyticities for physical $s>4m^2$ (assuming $\mathbb{Z}_2$ symmetry $\phi \ra -\phi$), as well as $t=-s-u+4m^2>4m^2$. The correct way to perform such an analytic continuation is to hold the spatial momenta of all particles fixed while deforming the energies of particles $2$ and $3$ to unphysical and possibly complex values (particles $2$ and $3$ do not need to be on-shell, but must have physical spatial momenta because we have inserted physical intermediate states).} $u \equiv 4 m^2 - s - t < 0$, the imaginary part of the amplitude comes from the poles in the $JJ$ terms (in the limit \(\epsilon \to 0\) recall \((x-i\epsilon)^{-1} = \text{P.V.}(x^{-1}) + i\pi \delta(x)\), where P.V. denotes the principal value),
\begin{equation} \label{eq:ImAmp}
\ba
    \rmIm[\Mcal(s,t)] &= \fr{\pi}{Z} \sum_\alpha \bigg( \<\ve{p}_4|J(0)|M_\alpha^2 ; \ve{p}_1 + \ve{p}_2 \>\<M_\alpha^2 ; \ve{p}_1 + \ve{p}_2 | J(0) |\ve{p}_1\> \, \delta(s - M_\alpha^2) \\
    & \hspace{2cm} + \<\ve{p}_4|J(0)|M_\alpha^2 ; \ve{p}_1 - \ve{p}_3 \>\<M_\alpha^2 ; \ve{p}_1 - \ve{p}_3 | J(0) |\ve{p}_1\> \, \delta(t - M_\alpha^2) \bigg).
\ea
\end{equation}
In other words, the $s$- and $t$-channel cuts manifestly come from intermediate on-shell states with mass $M_\alpha^2 = s,t$. The structure of Eq.~\eqref{eq:GeneralAmp} can therefore be understood as encoding the fixed-$u$ dispersion relation
\begin{equation} \label{eq:dispersion_relation}
    \Mcal(s,t) = \fr{1}{\pi} \int_0^\infty ds' \, \fr{\rmIm[\Mcal(s',t')]}{s'-s-i\epsilon} + \fr{1}{\pi} \int_0^\infty dt' \, \fr{\rmIm[\Mcal(s',t')]}{t'-t-i\epsilon} \, + \, \textrm{subtraction terms},
\end{equation}
where the ``subtraction terms'' are all contributions which do not vanish as $|s| \to \infty$ at fixed~$u$ and thus cannot be reconstructed from the imaginary part.  Interestingly, all such terms are contained within the $J'$ term, which is manifestly independent of $s$ and can be computed directly from the one-particle eigenstate.

\section{Example: \texorpdfstring{$O(N)$}{O(N)} model at large \texorpdfstring{$N$}{N}} \label{sec:ONmodel}

Now that we have a general procedure for obtaining nonperturbative scattering amplitudes, let's proceed to a fully-worked example: the strongly-coupled scalar $O(N)$ model at $N \to \infty$ in 2+1 dimensions.  As we briefly review below, the $2 \to 2$ scattering amplitude is calculable at large $N$, allowing us to directly compare our numerical results to exact expressions. However the technique presented here can be applied to far more general nonperturbative systems, where there is no existing method for computing the $S$-matrix.

The layout of this section is as follows.  In Sec.~\ref{sec:truncation_review} we give a brief review of lightcone conformal truncation (LCT), which we follow in Sec.~\ref{sec:ONtruncation} with a review of the $O(N)$ model at large $N$ (both analytically and within the context of truncation). Our aim here is to explain essential features and results of the calculation and not get bogged down in details; for those, we refer the reader to App.~\ref{app:ON_truncation_detail}. Finally, in Sec.~\ref{sec:truncation_results} we use Eq.~\eqref{eq:GeneralAmp} to calculate the nonperturbative $2 \to 2$ scattering amplitude with truncation and discuss various features of the result.

\subsection{Quick review of lightcone conformal truncation} \label{sec:truncation_review}

Hamiltonian truncation is a variational method for approximating the spectrum of a Hamiltonian. Such an approach is most familiar in nonrelativistic quantum mechanics, but there have been multiple truncation frameworks developed for applications to strongly-coupled QFTs. These methods, which differ in various details of their implementation, all share the same basic steps: (1) \emph{discretize} the Hilbert space of the QFT in some way, (2) \emph{truncate} the discretized Hilbert space to a finite-dimensional subspace, and (3) \emph{diagonalize} the Hamiltonian within this subspace, which is typically done numerically. The eigenstates of the truncated Hamiltonian provide a nonperturbative approximation to the low-energy eigenstates of the full QFT and can be used to compute general physical observables at arbitrary values of the couplings.

Lightcone conformal truncation~\cite{Katz:2016hxp} is a particular truncation method which studies general QFTs by viewing them as deformations of some UV conformal field theory (CFT) by one or more relevant operators $\Ocal_i$.
In this setup, the Hamiltonian takes the form
\begin{equation} \label{eq:CFTdeformation}
H_{\textrm{QFT}} = H_{\textrm{CFT}} + V, \qquad V = g_i \int d^{d-1} x \, \Ocal_i(x).
\end{equation}
Importantly, the couplings $g_i$ are \emph{not} necessarily taken to be small perturbations.

In any truncation approach, one must first choose a complete basis of states for the QFT Hilbert space. LCT uses the basis of momentum space states constructed from \emph{local operators} in the UV CFT:\footnote{Concretely, the basis is built from all primary operators, as descendants simply construct the same states with multiplicative factors of $p_\mu$: $|\p_\mu\Ocal; p_\mu\> = ip_\mu|\Ocal; p_\mu\>$.
}
\be
|\Ocal; p_\mu\> = \fr{1}{N_\Ocal} \int d^dx \, e^{-ip\cdot x} \Ocal(x)|0\>,
\label{eq:truncation_states_continuum}
\ee
where $N_\Ocal$ is  an overall normalization factor. Conceptually, these states are the CFT generalization of momentum space partial wave states, labeled by their overall momentum $p_\mu$ as well as the scaling dimension and spin quantum numbers of the operator $\Ocal$. In free CFTs, these correspond to multi-particle Fock space states weighted by polynomials in the individual particles' momenta.

Because we are interested in deformations of the form Eq.~\eqref{eq:CFTdeformation} (a spatial integral over a local operator) which do not violate translation invariance, we can work in a sector of fixed spatial momentum $\ve{p}$.  However, each CFT operator still creates a continuum of partial wave states $|\Ocal;p_\mu\>$ parametrized by the invariant mass $\mu^2 \equiv p^2$. This continuum of states must be discretized, which is done by introducing \emph{smearing functions} $b_\ifrak(\mu)$:
\be \label{eq:truncation_states}
|\Ocal; \ifrak; \ve{p} \> = \int_0^{\LambdaUV^2} d \mu^2 \, b_\ifrak(\mu) |\Ocal; p_\mu\>.
\ee
The resulting basis states are therefore labeled by the scaling dimension and spin of the local operator $\Ocal$, the spatial momentum $\ve{p}$, and the discrete label $\ifrak$ associated with the smearing function. Any finite set of smearing functions will necessarily introduce an effective UV and IR cutoff.  Here we make the UV cutoff manifest as $\LambdaUV$, which is taken to be much larger than the scales associated with any dimensionful couplings $g_i$. The smearing functions can in principle take any form, but in this work we use non-overlapping window functions binned in $\mu^2$.  For more details on the precise functions chosen, see App.~\ref{app:ON_truncation_detail}.

Once we have discretized the Hilbert space, we truncate the number of CFT states by restricting to operators with scaling dimension $\De \leq \Dmax$ (in free CFTs this simply corresponds to restricting to momentum space polynomials up to a certain degree), and we truncate the number of smearing functions for each CFT partial wave with the parameter $\imax$ (i.e.~we use $\imax$ bins in $\mu^2$ for each operator $\Ocal$). We are therefore left with a finite-dimensional subspace controlled by two parameters: $\Dmax$ and $\imax$.

Next, we evaluate the matrix elements of the full QFT Hamiltonian within this truncated basis of states and diagonalize the resulting finite-dimensional matrix. We thus obtain the eigenstates of the truncated Hamiltonian, expressed in terms of the CFT basis:
\be
|M_\alpha^2; \ve{p}\> = \sum_{\substack{\De \leq \Dmax \\ \ifrak \, \leq \, \imax}} C^{\Ocal;\ifrak}_\alpha |\Ocal; \ifrak; \ve{p} \>.
\ee

In principle, this truncation scheme can be implemented in any quantization scheme. In this work, we use lightcone quantization, with the Hilbert space defined on slices of fixed ``lightcone time'' $x^+ \equiv \fr{1}{\sqrt{2}}(x^0 + x^1)$. This choice of quantization scheme allows us to work directly in infinite volume, which is most natural for studying the $S$-matrix. More details on the exact formulation of this approach for the example of the $O(N)$ model can be found in App.~\ref{app:ON_truncation_detail}.

\subsection{\texorpdfstring{$O(N)$ model scattering amplitude in truncation}{O(N) model scattering amplitude in truncation}} \label{sec:ONtruncation}

We now consider a theory consisting of $N$ massive scalar fields $\phi^i$ ($i = 1, ..., N$) in $d = 2+1$ dimensions with the Lagrangian
\begin{equation} \label{eq:ONlagrangian}
    \Lcal = \half \del^\mu \phi^i \del_\mu \phi^i - \half m^2 \phi^i \phi^i - \frac{\lambda}{4 N} (\phi^i \phi^i) (\phi^j \phi^j).
\end{equation}
In the limit $N \to \infty$, this theory is solvable for all values of $m$ and $\lambda$, making it a useful testing ground for our truncation technique at strong coupling.  Here we will focus on the scattering amplitude for $\phi^i \phi^j \to \phi^k \phi^\ell$, which at leading order in $1/N$ can be computed analytically at arbitrary coupling $\lambda$ by summing the bubble diagrams in Fig.~\ref{fig:ON_amp_diagram}:
\begin{equation} \label{eq:amp_ON_all_flavors}
    \Mcal^{i j k \ell}(s, t) = \fr{1}{N} \Big( \Mcal(s) \delta^{i j} \delta^{k \ell} + \Mcal(t) \delta^{i k} \delta^{j \ell} + \Mcal(u) \delta^{i \ell} \delta^{j k} \Big) + \order\bigg( \fr{1}{N^2} \bigg),
\end{equation}
with the flavor-stripped amplitude given by~\cite{Moshe:2003xn}
\begin{equation} \label{eq:amp_single_flavor}
    \Mcal(s) = - \frac{
        2\lambda
        }{
        1 + \frac{\lambda}{8 \pi \sqrt{s}} \left[ \log \left( \frac{\sqrt{s} + 2 m}{\sqrt{s} - 2 m} \right) + i \pi \right]
        }.
\end{equation}
We will refer to the three distinct contributions in~\eqref{eq:amp_ON_all_flavors} as the $s$-flavor, $t$-flavor, and $u$-flavor amplitudes because of their $O(N)$ flavor index structure.

\begin{figure}
    \centering
    \includegraphics[width=0.9\linewidth]{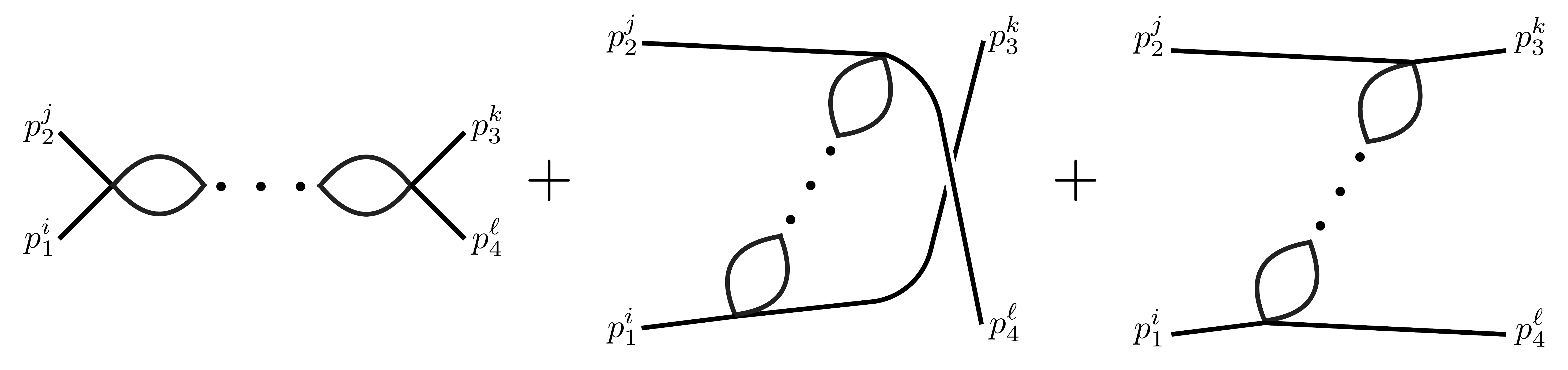}
    \caption{The $O(N)$ model amplitude at large $N$, Eq.~\eqref{eq:amp_ON_all_flavors}, is obtained from resumming all exchanged bubble chains in the $s$-channel (left), $t$-channel (middle), and $u$-channel (right).}
    \label{fig:ON_amp_diagram}
\end{figure}

Our task now is to reproduce the above expression using truncation, although we will take the $N \to \infty$ limit analytically to keep the example simple.\footnote{
Concretely, we will exploit the fact that certain Hamiltonian matrix elements vanish in the infinite $N$ limit and that most basis states decouple from the relevant dynamics, which reduces the numerical complexity of the problem. We leave a study of finite-$N$ effects for future work.} Many details of the application of LCT to the $O(N)$ model have been previously discussed in the literature (see e.g.~\cite{Katz:2016hxp,Delacretaz:2018xbn}), so here we will only briefly review the relevant prior results and instead focus on how they are used to compute the amplitude with the recipe presented in Sec.~\ref{sec:Smat_from_estates}.

We first split the Lagrangian into that of a free CFT and two deformations,
\begin{equation}
    \Lcal = \Lcal^{\text{(CFT)}} + \Lcal^{(m)} + \Lcal^{(\lambda)},
\end{equation}
where $\Lcal^{\text{(CFT)}} \equiv \half (\p\vec{\phi} \,)^2$, $\Lcal^{(m)} \equiv - \half m^2 \vec{\phi}^{\,2}$, and $\Lcal^{(\lambda)} \equiv - \frac{\lambda}{4 N} (\vec{\phi}^{\,2})^2$. We can now follow the general procedure outlined in Sec.~\ref{sec:truncation_review} by constructing a basis of states built from operators in the free CFT with $\De \leq \Dmax$. These operators take the schematic form
\be
\Ocal^{i_1 \cdots i_n} = \sum_{k_1,\ldots,k_n} C^\Ocal_{k_1,\ldots,k_n} \p^{k_1} \phi^{i_1} \cdots \p^{k_n} \phi^{i_n},
\ee
and can be organized by particle number, scaling dimension, spin, and $O(N)$ representation. We can then compute the matrix elements of the full Hamiltonian (CFT + deformations) within this truncated basis of states. Note that particle-number-changing matrix elements are suppressed by $1/N$~\cite{Katz:2016hxp}, which simplifies the structure of the Hamiltonian and its eigenstates in the large $N$ limit.

Once we have diagonalized the truncated Hamiltonian to obtain the approximate energy eigenstates $|M_\alpha^2;\ve{p}\>$, we can compute the scattering amplitude. Due to the presence of $N$ scalar fields, we must first generalize Eq.~\eqref{eq:LSZinTermsOfJJ} to include flavor indices, obtaining the  expression
\be
\ba
    \<\onmom{3}{k},\onmom{4}{\ell}|\onmom{1}{i},\onmom{2}{j}\> &= \left( \frac{i}{\sqrt{Z}} \right)^2 \int d^3 x_2 \, d^3 x_3 \, e^{i (p_3 \cdot x_3 - p_2 \cdot x_2)} \\
    &\qquad \qquad \times 
    \bigg( \< \onmom{4}{\ell} | T\{ J^k (x_3) J^j (x_2) \}| \onmom{1}{i} \> - i \delta^3(x_3 - x_2) \< \onmom{4}{\ell} | J'^{\,jk} (x_2) | \onmom{1}{i} \> \bigg),
\ea
\ee
where $| \onmom{m}{i} \>$ is the one-particle energy eigenstate with momentum $\ve{p}_m$ and flavor index $i$, and
\be
J^i \equiv - \frac{\lambda}{N} \vec{\phi}^{\,2} \phi^i, \qquad J'^{\,ij} \equiv \frac{\del J^i}{\del \phi^j} = \frac{\del J^j}{\del \phi^i} = - \fr{\lambda}{N} \big( 2 \phi^i \phi^j + \delta^{i j} \vec{\phi}^{\,2} \big).
\ee

Inserting the approximate energy eigenstates, as in Sec.~\ref{sec:Smat_from_estates}, we arrive at the final result
\be \label{eq:amp_arbitrary_flavor}
\ba
    \Mcal^{i j k \ell}(s,t) &= \frac{1}{Z} \Bigg[ \sum_{\alpha} \bigg( \frac{1}{M_\alpha^2 - s - i \epsilon} \< \onmom{4}{\ell} | J^k | M_\alpha^2 ; \ve{p}_1 + \ve{p}_2 \> \< M_\alpha^2 ; \ve{p}_1 + \ve{p}_2 | J^j | \onmom{1}{i} \> \\
    & \quad + \frac{1}{M_\alpha^2 - t - i \epsilon} \< \onmom{4}{\ell} | J^j | M_\alpha^2 ; \ve{p}_1 - \ve{p}_3 \> \< M_\alpha^2 ; \ve{p}_1 - \ve{p}_3 | J^k | \onmom{1}{i} \> \bigg) + \< \onmom{4}{\ell} | J'^{\,jk} | \onmom{1}{i} \> \Bigg].
\ea
\ee
We refer to the two different $JJ$ contributions in parentheses as the ``$s$-kinematics'' and ``$t$-kinematics'' terms because of their form as sums over poles in $s$ and $t$, respectively. We will now use Eq.~\eqref{eq:amp_arbitrary_flavor} to reconstruct all three contributions to the leading $1/N$ scattering amplitude~\eqref{eq:amp_ON_all_flavors}.

\subsubsection*{\texorpdfstring{$s$}{s}-flavor amplitude}

Let's first consider the $s$-flavor amplitude, represented schematically by the left bubble chain diagram in Fig.~\ref{fig:ON_amp_diagram}, which can be obtained from~\eqref{eq:amp_arbitrary_flavor} via
\be
\Mcal(s) \equiv \lim_{N \ra \infty} \fr{1}{N} \Mcal^{iijj}(s,t),
\ee
where repeated indices are summed. At finite $N$, or in a generic nonperturbative theory, we would insert \emph{all} approximate eigenstates $|M_\alpha^2; \ve{p}\>$ in \eqref{eq:amp_arbitrary_flavor} to compute this amplitude. At large $N$, however, the contributions of most intermediate states are suppressed by higher powers of $1/N$, which simplifies the calculation.
In particular, the corrections to the external one-particle states are suppressed at large $N$, such that $|\ve{p}_1^i\>$ and $\<\ve{p}_4^\ell|$ simply correspond to \emph{free} one-particle states and the field strength factor $Z=1 + \order(1/N)$. Furthermore, for physical external momenta (namely, $p_{i-} > 0$) the only unsuppressed contribution to $\Mcal(s)$ comes from two-particle flavor-singlet intermediate states in the ``$s$-kinematics'' $JJ$ term (plus the contact interaction from the $J'$ term), as we demonstrate in App.~\ref{app:ON_truncation_detail}.

\begin{figure}
    \centering
    \includegraphics[width=0.9\linewidth]{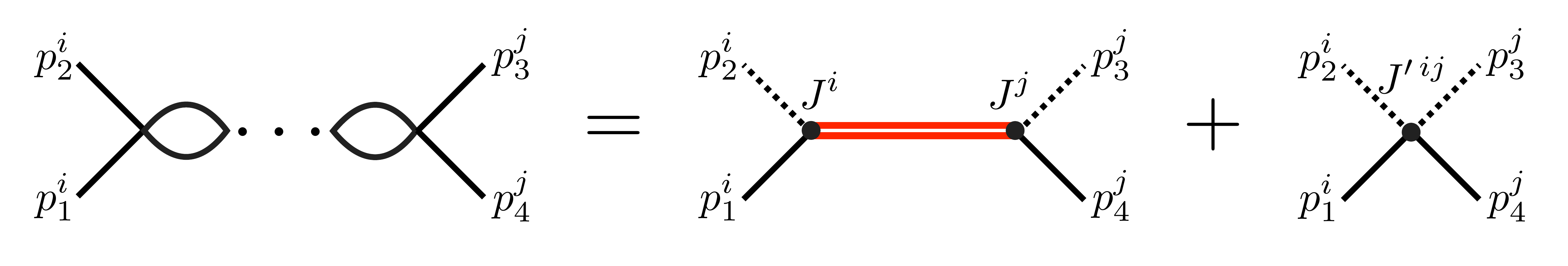}
    \caption{The $s$-flavor scattering amplitude (left) is obtained in truncation from the intermediate propagation of two-particle flavor-singlet energy eigenstates (middle) and the $J'$ contact interaction~(right).}
    \label{fig:schannel_amp_truncation}
\end{figure}

We therefore find the simple result as $N\to\infty$,
\be
    \fr{1}{N} \Bigg( \! \sum_{\alpha \in \{n=2\}} \! \frac{{}_{\textrm{free}}\< \onmom{4}{j} | J^j | M_{\alpha}^2 ; \ve{p}_1 + \ve{p}_2 \> \< M_\alpha^2 ; \ve{p}_1 + \ve{p}_2 | J^i | \onmom{1}{i} \>_{\textrm{free}}}{M_\alpha^2 - s - i \epsilon} + {}_{\textrm{free}}\< \onmom{4}{j} | J'^{\,ij} | \onmom{1}{i} \>_{\textrm{free}}\! \Bigg) \rightarrow \, \Mcal(s),
    \label{eq:schannel_truncation}
\ee
where the sum is specifically over two-particle intermediate eigenstates and $| \onmom{1}{i} \>_{\textrm{free}}$ indicates that the external one-particle states are those of the free theory ($\lambda=0$, $m\neq0$). This result is shown schematically in Fig.~\ref{fig:schannel_amp_truncation}. On the RHS of this diagrammatic equation, solid black lines represent free external particles, dashed black lines represent external propagators which have been explicitly canceled using the equations of motion, and double red lines represent intermediate two-particle interacting energy eigenstates obtained by diagonalizing the truncated Hamiltonian.

Truncation therefore directly reproduces our physical intuition from Feynman diagrams: the $s$-flavor sum of bubble chains at large $N$ comes from two-particle intermediate states. In Sec.~\ref{sec:truncation_results}, we confirm Eq.~\eqref{eq:schannel_truncation} by comparing the results obtained with truncation to the exact analytic expression~\eqref{eq:amp_single_flavor}.

\subsubsection*{\texorpdfstring{$t$}{t}-flavor amplitude}

Let's now turn to the $t$-flavor amplitude, corresponding to the middle bubble chain diagram in Fig.~\ref{fig:ON_amp_diagram},
\be
\Mcal(t) \equiv \lim_{N \ra \infty} \fr{1}{N} \Mcal^{ijij}(s,t).
\ee
In truncation, the structure of the $t$-flavor amplitude is conceptually very similar to the $s$-flavor one. However for physical external momenta this amplitude naively receives contributions from both the $s$- and $t$-kinematics $JJ$ terms. As we show in App.~\ref{app:ON_truncation_detail}, though, the $s$-kinematics term is only nonzero for $p_{1-} < p_{3-}$, while the $t$-kinematics term is only nonzero for $p_{1-} > p_{3-}$. For a given choice of external momenta, this amplitude therefore only receives a contribution from one of the two $JJ$ terms. For simplicity, here we'll focus on the case $p_{1-} > p_{3-}$, leaving the other for the appendix.

\begin{figure}
    \centering
    \includegraphics[width=0.9\linewidth]{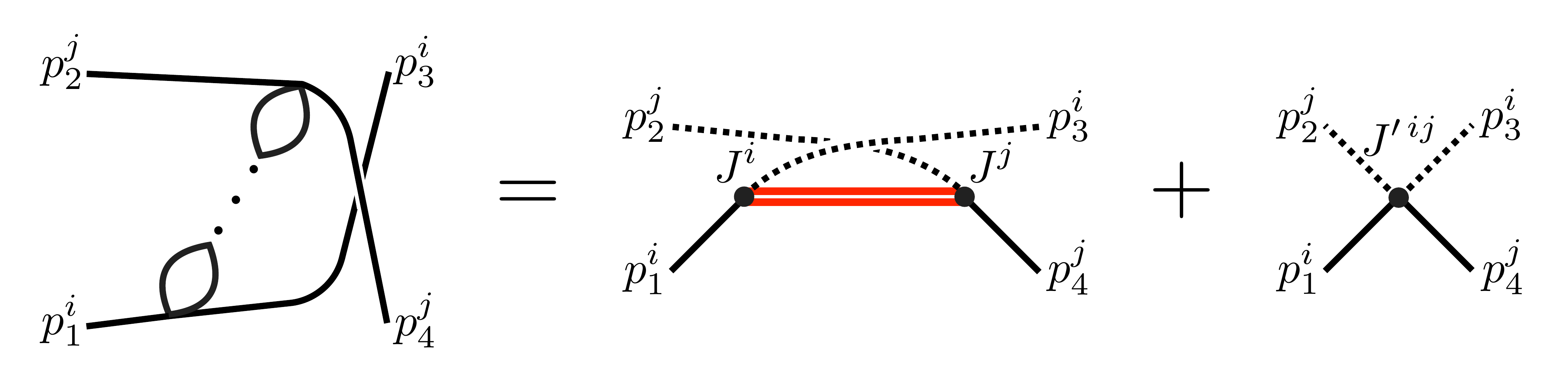}
    \caption{The $t$-flavor scattering amplitude (left) is obtained in truncation from the intermediate propagation of two-particle flavor-singlet energy eigenstates (middle) and the $J'$ contact interaction (right). Note that this diagram specifically describes the case where $p_{1-} > p_{3-}$, as the intermediate two-particle states must have physical momentum. This expression exactly matches the $s$-flavor scattering amplitude (Fig.~\ref{fig:schannel_amp_truncation}) with $p_2 \lra - p_3$.}
    \label{fig:tchannel_amp_truncation}
\end{figure}

In this case, the only unsuppressed contribution to $\Mcal(t)$ (other than the contact interaction in the $J'$ term) again comes from two-particle flavor-singlet intermediate states, now in the ``$t$-kinematics'' $JJ$ term. Unsurprisingly, due to the manifest $s \leftrightarrow t$ crossing symmetry in Eq.~\eqref{eq:amp_arbitrary_flavor}, the $t$-flavor amplitude thus corresponds to the $s$-flavor one with $\onmom{2}{j} \leftrightarrow -\onmom{3}{k}$,
\be
\fr{1}{N} \Bigg( \! \sum_{\alpha \in \{n=2\}} \! \frac{{}_{\textrm{free}}\< \onmom{4}{j} | J^j | M_{\alpha}^2 ; \ve{p}_1 - \ve{p}_3 \> \< M_\alpha^2 ; \ve{p}_1 - \ve{p}_3 | J^i | \onmom{1}{i} \>_{\textrm{free}}}{M_\alpha^2 - t - i \epsilon} + {}_{\textrm{free}}\< \onmom{4}{j} | J'^{\,ij} | \onmom{1}{i} \>_{\textrm{free}}\! \Bigg) \rightarrow \, \Mcal(t),
    \label{eq:tchannel_truncation}
\ee
represented schematically in Fig.~\ref{fig:tchannel_amp_truncation}. As discussed in App.~\ref{app:ON_truncation_detail}, the other case ($p_{1-} < p_{3-}$) instead arises from the exchange of four-particle intermediate states in the $s$-kinematics term, but this expression can be rewritten into a form similar to Eq.~\eqref{eq:tchannel_truncation}.

\subsubsection*{\texorpdfstring{$u$}{u}-flavor amplitude}

Finally, we have the $u$-flavor amplitude, corresponding to the right bubble chain diagram in Fig.~\ref{fig:ON_amp_diagram},
\be
\Mcal(u) \equiv \lim_{N \ra \infty} \fr{1}{N} \Mcal^{ijji}(s,t).
\ee
In our setup---where we have made the arbitrary choice to reduce particles 2 and 3---the structure of this amplitude is fundamentally distinct from that of the $s$- and $t$-flavor amplitudes. At large $N$, it actually receives \emph{no contribution} from the $JJ$ term in~\eqref{eq:amp_arbitrary_flavor}, and arises solely from the $J'$ term.

Naively, $\< \onmom{4}{i} | J'^{\,jj} | \onmom{1}{i} \>$ is simply an $\order(\lambda)$ contact term, as all corrections to the external one-particle states are suppressed by higher powers of $1/N$. However, the contractions of $J'$ with itself and the two external particles with each other give an additional factor of $N$ to the matrix elements for these corrections, which exactly compensates for this suppression, such that the $u$-flavor amplitude receives a contribution from the mixing of the one-particle state with three-particle states in the same flavor representation. As we show in App.~\ref{app:ON_truncation_detail}, at large $N$ these three-particle states can be written as the tensor product of 
two-particle flavor-singlet states and a free one-particle state, resulting in the leading correction\footnote{Note that here $E$ refers to the lightcone energy $P_+$, see App.~\ref{app:ON_truncation_detail} for details.}
\be
| \onmom{1}{i} \> = | \onmom{1}{i} \>_{\textrm{free}} + \!\!\! \sum_{\alpha\in\{n=2\}} \! \int_{\ve{q}} \fr{{}_{\textrm{free}} \<\ve{q}^j| \otimes \<M_\alpha^2; \ve{p}_1 - \ve{q}|V^{(\lambda)}| \onmom{1}{i} \>_{\textrm{free}}}{E_1 - (E_\alpha + E_4)} \, |M_\alpha^2;\ve{p}_1 - \ve{q}\> \otimes |\ve{q}^j\>_{\textrm{free}} + \order\bigg( \fr{1}{N^2} \bigg),
\ee
where $V^{(\lambda)}$ is the potential due to the quartic interaction,
\be
V^{(\lambda)} = \fr{\lambda}{4N} \int d^2x \, (\vec{\phi}^{\,2})^2.
\ee

\begin{figure}
    \centering
    \includegraphics[width=0.9\linewidth]{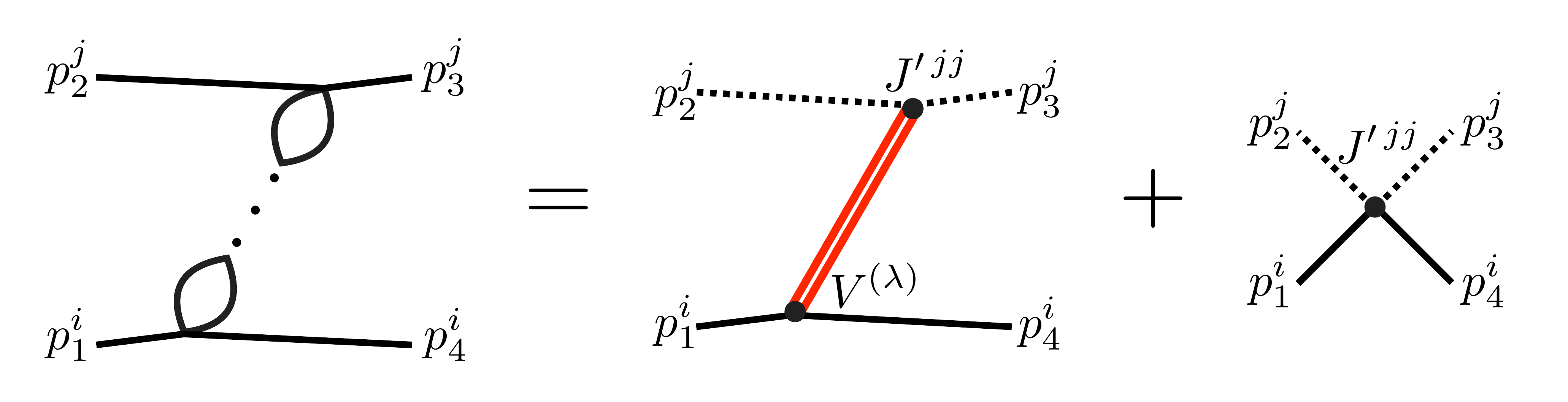}
    \caption{The $u$-flavor scattering amplitude (left) is obtained in truncation from the free $J'$ contact interaction (right) and the mixing of external one-particle states with intermediate three-particle states (which are tensor products of flavor-singlet two-particle states with a free one-particle state) in the $J'$ matrix elements (middle). Note that this diagram specifically describes the case where $p_{1-} > p_{4-}$, as the intermediate two-particle states must have physical momentum.}
    \label{fig:uchannel_amp_truncation}
\end{figure}

Expanding $\< \onmom{4}{i} | J'^{\,jj} | \onmom{1}{i} \>$ to leading order in $1/N$, we obtain an analogous structure to the $t$-flavor amplitude, where the correction to $| \onmom{1}{i} \>$ only gives a nonzero contribution for $p_{1-} > p_{4-}$, while the correction to $\< \onmom{4}{i} |$ gives a nonzero contribution for $p_{1-} < p_{4-}$. Let's focus on the case $p_{1-} > p_{4-}$, where we find the result:
\be
\ba
    &\fr{1}{N} \Bigg( \! \sum_{\alpha\in\{n=2\}} \frac{\big({}_{\textrm{free}}\< \onmom{4}{i} | J'^{\,jj} | M_\alpha^2 ; \ve{p}_1 - \ve{p}_4 \> \otimes |\onmom{4}{k}\>_{\textrm{free}} \big) \big({}_{\textrm{free}} \<\onmom{4}{k}| \otimes \< M_\alpha^2 ; \ve{p}_1 - \ve{p}_4 | V^{(\lambda)} | \onmom{1}{i} \>_{\textrm{free}}\big)}{E_1 - (E_\alpha + E_4)} \\
    & \hspace{2.5cm} + {}_{\textrm{free}}\< \onmom{4}{i} | J'^{\,jj} | \onmom{1}{i} \>_{\textrm{free}} \! \Bigg) \ra \Mcal(u),
    \label{eq:uchannel_truncation}
\ea
\ee
shown schematically in Fig.~\ref{fig:uchannel_amp_truncation}. As we demonstrate in the appendix, this expression can be rewritten into a similar form to Eqs.~\eqref{eq:schannel_truncation} and \eqref{eq:tchannel_truncation}.

\subsection{Results} \label{sec:truncation_results}

With the above results assembled, we now explicitly demonstrate the computation of $S$-matrix elements with Hamiltonian truncation. Following the procedure laid out in Sec.~\ref{sec:truncation_review}, we first construct a basis of states, whose size is controlled by the truncation parameters $\Dmax$ and $\imax$, and evaluate the Hamiltonian matrix elements in this truncated basis (see App.~\ref{app:ON_truncation_detail} for a detailed presentation). As we increase these two parameters, the size of the basis grows and the eigenstates of the truncated Hamiltonian approach the exact QFT energy eigenstates. We then follow the approach of Sec.~\ref{sec:ONtruncation} and insert the resulting eigenstates into Eq.~\eqref{eq:amp_arbitrary_flavor} to obtain the $\phi^i \phi^j \ra \phi^k \phi^\ell$ scattering amplitude.

For concreteness, let's focus on the $s$-flavor scattering amplitude $\Mcal(s)$. From Eq.~\eqref{eq:schannel_truncation}, we see that we only need the flavor-singlet two-particle energy eigenstates to compute this observable. As shown in App.~\ref{app:ON_truncation_detail}, the matrix elements of $J$ and $J'$ in~\eqref{eq:schannel_truncation} are manifestly real, so for real values of $s$ the imaginary part of the scattering amplitude comes solely from the denominator
\begin{equation}\label{eq:finiteeps}
\frac{1}{M_\alpha^2 - s-i\epsilon}=\frac{M_\alpha^2 - s}{(M_\alpha^2 - s)^2+\epsilon^2}+i\frac{\epsilon}{(M_\alpha^2 - s)^2+\epsilon^2}\,\,\xrightarrow[\epsilon\to0]{}\,\,\textrm{P.V.} \frac{1}{M_\alpha^2 - s}+i\pi \delta(s - M_\alpha^2),
\end{equation}
where $\PV$ indicates the principal value. In the limit of infinitesimal $\epsilon$, we thus obtain\footnote{The overlaps $\< \vec{\phi}^{\,2}| M_\alpha^2\>$ arise from the matrix elements of $J^i \sim \vec{\phi}^{\,2} \phi^i$ (see App.~\ref{app:ON_truncation_detail}). These overlaps are $\order(\sqrt{N})$, such that the resulting amplitude is finite as $N \ra \infty$.}
\begin{eqnarray} \label{eq:s_kinematics_imag}
    \rmIm [\Mcal(s)] &=& \sum_\alpha \pi \delta(s - M_\alpha^2) \fr{\lambda^2}{N} \Big| \< \vec{\phi}^{\,2}(0) | M_\alpha^2 ; \ve{p}_1 + \ve{p}_2 \> \Big|^2, \\
\label{eq:s_kinematics_real}
    \rmRe [\Mcal(s)] &=& - 2 \lambda + \sum_\alpha \PV \frac{1}{M_\alpha^2 - s} \frac{\lambda^2}{N} \Big| \< \vec{\phi}^{\,2}(0) | M_\alpha^2 ; \ve{p}_1 + \ve{p}_2 \> \Big|^2,
\end{eqnarray}
where the $-2\lambda$ in the real part arises from the $J'$ term (see App.~\ref{app:ON_truncation_detail}). These expressions make clear a general feature of the construction of the $S$-matrix from energy eigenstates: at a fixed (physical) value of $s$, the imaginary part of the amplitude is sensitive only to those states with invariant mass $M_\alpha^2 \approx s$, while the real part is sensitive to all states in the spectrum (albeit with increased sensitivity to those states with $M_\alpha^2 \approx s$).  For this reason we generically expect the truncation results for the real and imaginary parts of an amplitude to have different convergence properties as we increase the size of the basis.

Because the truncation spectrum is a discrete approximation to the QFT continuum, the imaginary part of the amplitude~\eqref{eq:s_kinematics_imag} appears as a sum of delta functions while the true amplitude is a smooth function of $s$.\footnote{The integrated imaginary part $\mathcal{I}(s) \equiv \int_{4 m^2}^{s} ds' \, \rmIm [\Mcal (s')] $ obtained from truncation is a much smoother function, analogously to the integrated spectral densities computed in e.g.\ Ref.~\cite{Katz:2016hxp}.} In practice, we therefore need to ``smear'' the imaginary part over a small but finite range $\xi s$,
\begin{equation} \label{eq:smeared_im_part}
    \rmIm [\Mcal(s)]_\xi \equiv \frac{1}{2 \xi s} \int_{s (1 - \xi)}^{s (1 + \xi)} ds' \, \rmIm [\Mcal (s')]. 
\end{equation}
This smearing procedure is conceptually similar to taking finite $\epsilon$ in~\eqref{eq:finiteeps}.\footnote{
In a system with discretized energy, it is natural to think of its analytic continuation as involving finite resolution also in the imaginary direction. Then,
a finite $\epsilon$ offers a practical way of implementing  the Feynman $i\epsilon$ prescription: propagators lead to smooth peaked functions rather than sharp poles, see the central expression in \eqref{eq:finiteeps}.} Provided the scale $\xi s$ is much smaller than the scale of any physical features (thresholds, resonances, etc.) in the amplitude, this will yield an accurate approximation to the true result.

\begin{figure}
    \centering
    \includegraphics[width=0.45\linewidth]{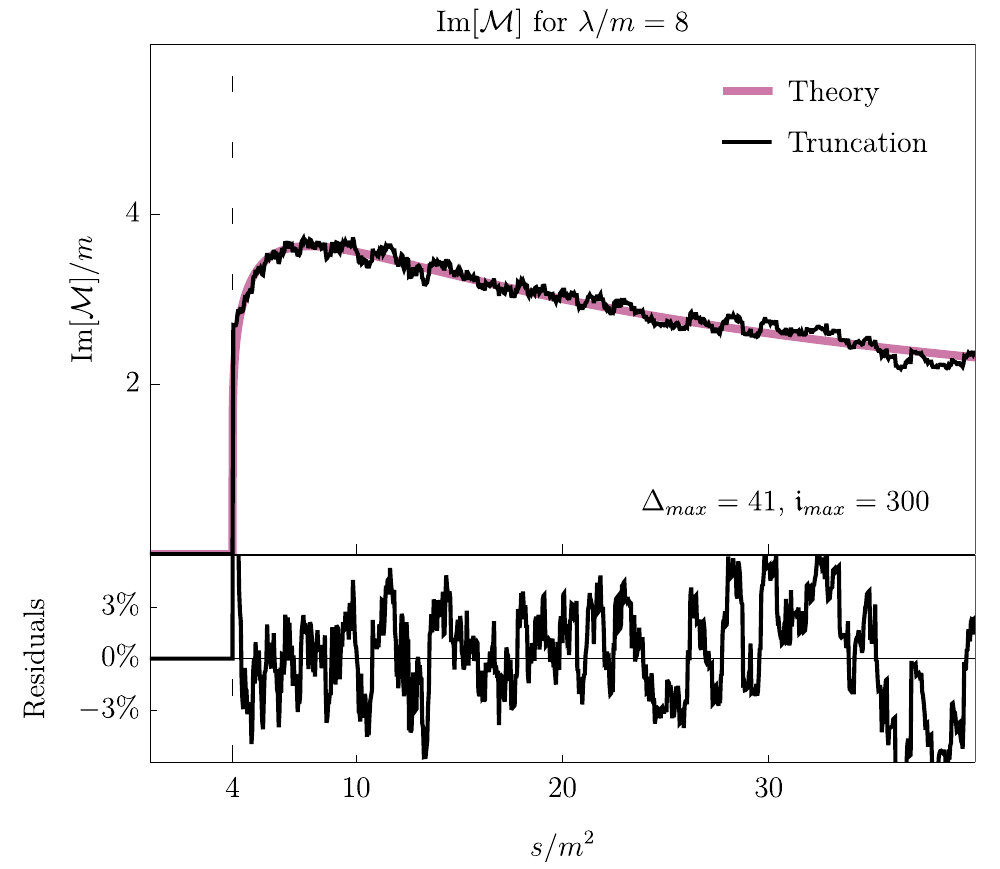}
    \includegraphics[width=0.45\linewidth]{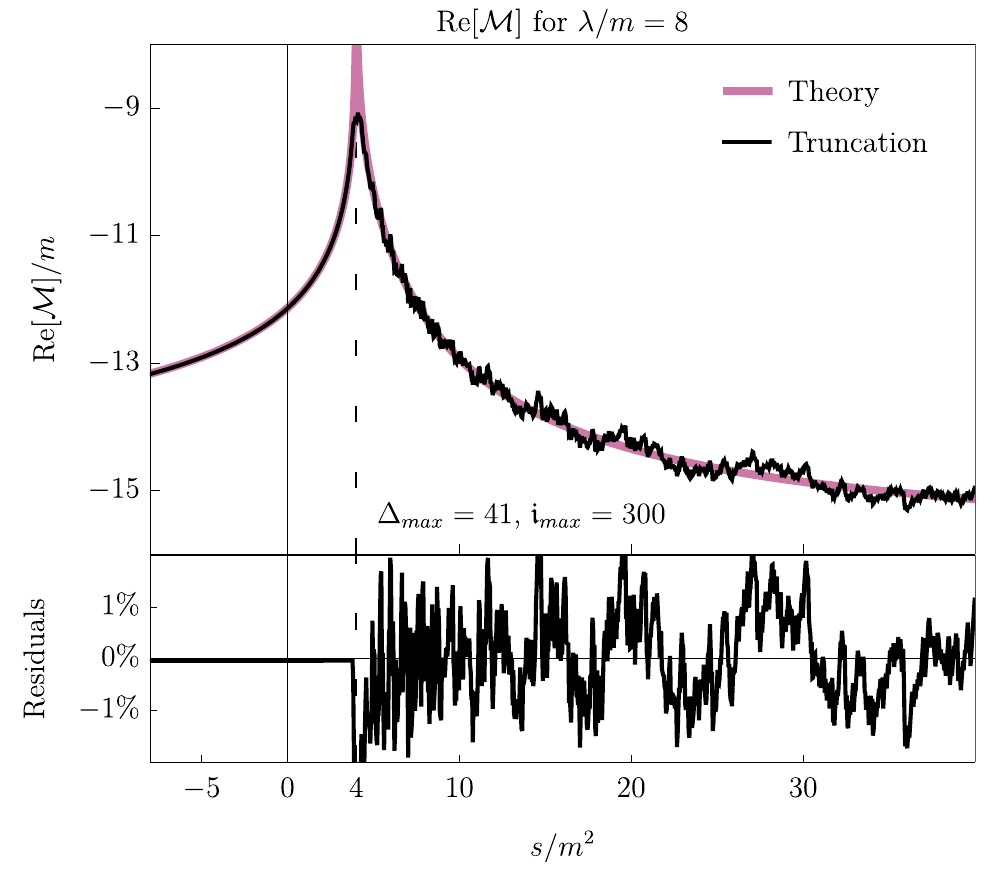}
    \caption{\footnotesize{\textbf{Left:} Comparison of the truncation and theoretical results for the imaginary part of the $s$-flavor amplitude.  \textbf{Right:} The same comparison for the real part.  Both of these plots are made with truncation parameters $\Dmax = 41$, $\imax = 300$ (6020 states), and the truncation calculations use the smearing procedures of Eqs.~\eqref{eq:smeared_im_part} and~\eqref{eq:cutoff_re_part}, respectively, with $\xi = 0.15$ for the imaginary part and $\xi = 0.05$ for the real part (because the real and imaginary part have different convergence properties it is best to choose different values of $\xi$ for each). The dotted line in each plot is at the threshold $s = 4 m^2$.}}
    \label{fig:sChanResiduals}
\end{figure}

The left plot in Fig.~\ref{fig:sChanResiduals} shows the imaginary part of $\Mcal(s)$ at strong coupling ($\fr{\lambda}{m} = 8$) obtained using Eq.~\eqref{eq:s_kinematics_imag}, with $\Dmax=41$, $\imax = 300$ (a total of 6020 basis states), and $\xi = 0.15$. As we can see, the truncation results (black) reproduce the known exact expression (magenta) from Eq.~\eqref{eq:amp_single_flavor} to percent-level accuracy over a wide range of energies.

Turning to the real part of the amplitude, the truncation result~\eqref{eq:s_kinematics_real} suffers from a similar problem: states with $M_\alpha^2 \approx s$ give large contributions to the amplitude, but with opposite signs depending on whether $M_\alpha^2 < s$ or $M_\alpha^2 > s$. In the true continuum theory, states on either side of $s$ give contributions that almost perfectly cancel each other, but when the system is discretized via truncation this cancellation is lost.

Taking inspiration from the finite $\epsilon$ version of~\eqref{eq:finiteeps}, where the contribution from states with $M_\alpha^2$ close to $s$ is damped, we choose to ``smear'' the real part by removing states within a small window $\xi s$, i.e.~summing only over states with $|M_\alpha^2 - s| > \xi s$,
\begin{equation} \label{eq:cutoff_re_part}
     \rmRe[ \Mcal(s)]_\xi \equiv - 2 \lambda + \sum_{|M_\alpha^2 - s| > \xi s} \PV \frac{1}{M_\alpha^2 - s} \frac{\lambda^2}{N^2} \Big| \< \vec{\phi}^{\,2}(0) | M_\alpha^2 ; \ve{p}_1 + \ve{p}_2 \> \Big|^2.
\end{equation}
The right plot in Fig.~\ref{fig:sChanResiduals} shows the resulting real part of $\Mcal(s)$ with $\xi = 0.05$ (black), which again reproduces the exact expression (magenta) to the percent-level.

Note that Eq.~\eqref{eq:schannel_truncation} is not limited to physical $s > 4m^2$, but can actually be evaluated for generic $s \in \mathbb{C}$, allowing us to study the analytic structure of the amplitude. In Fig.~\ref{fig:complexSresiduals} we demonstrate this explicitly by comparing the truncation results ($\Dmax = 41$, $\imax=300$) for the magnitude (left) and phase (right) of $\Mcal(s)$ to the analytic expression with complex~$s$. As we can see, the convergence of the truncation results is even better away from the physical regime, with sub-percent-level accuracy across most of the complex plane.

\begin{figure}
    \centering
    \includegraphics[width=0.48\linewidth]{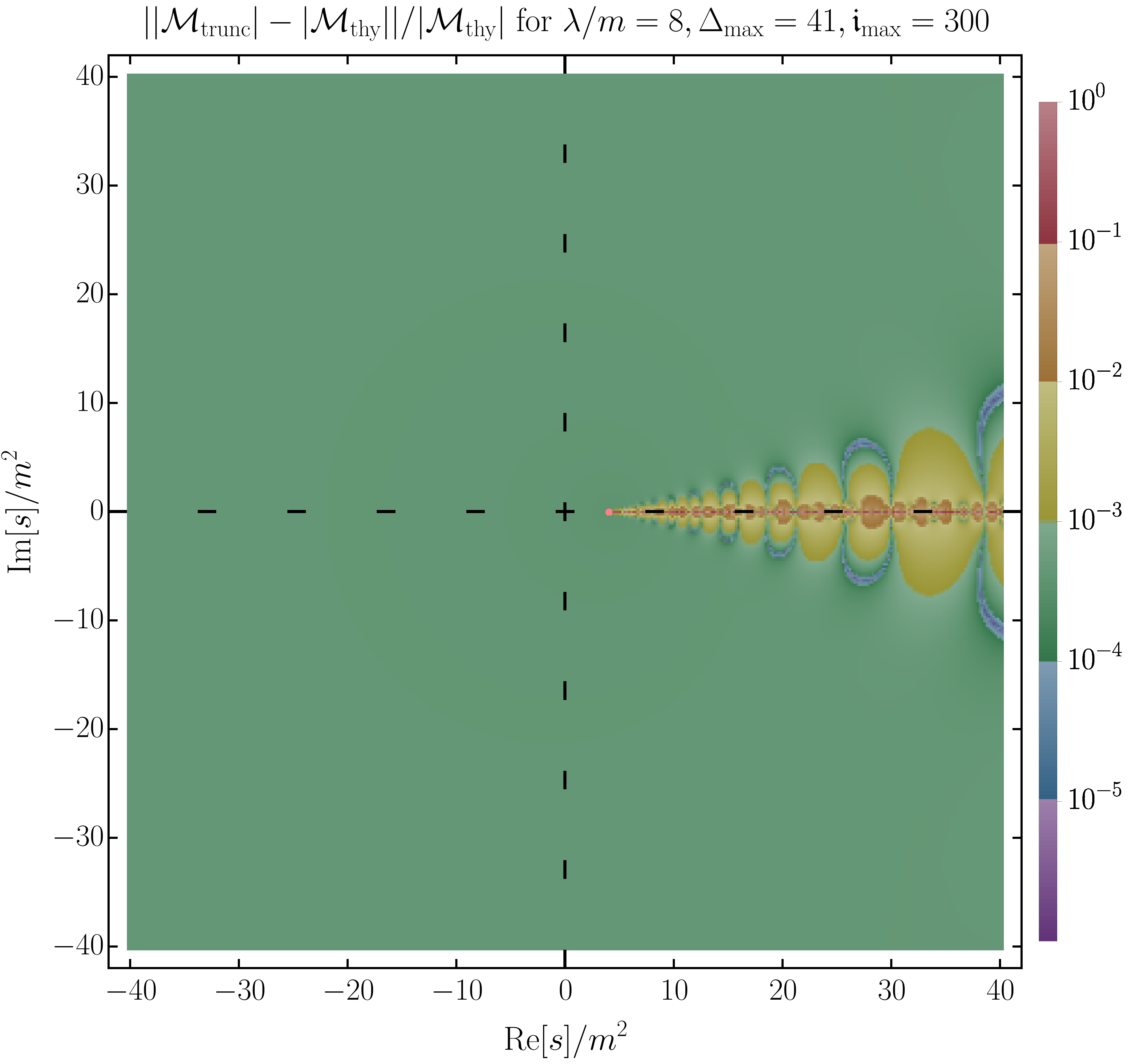}
    \includegraphics[width=0.48\linewidth]{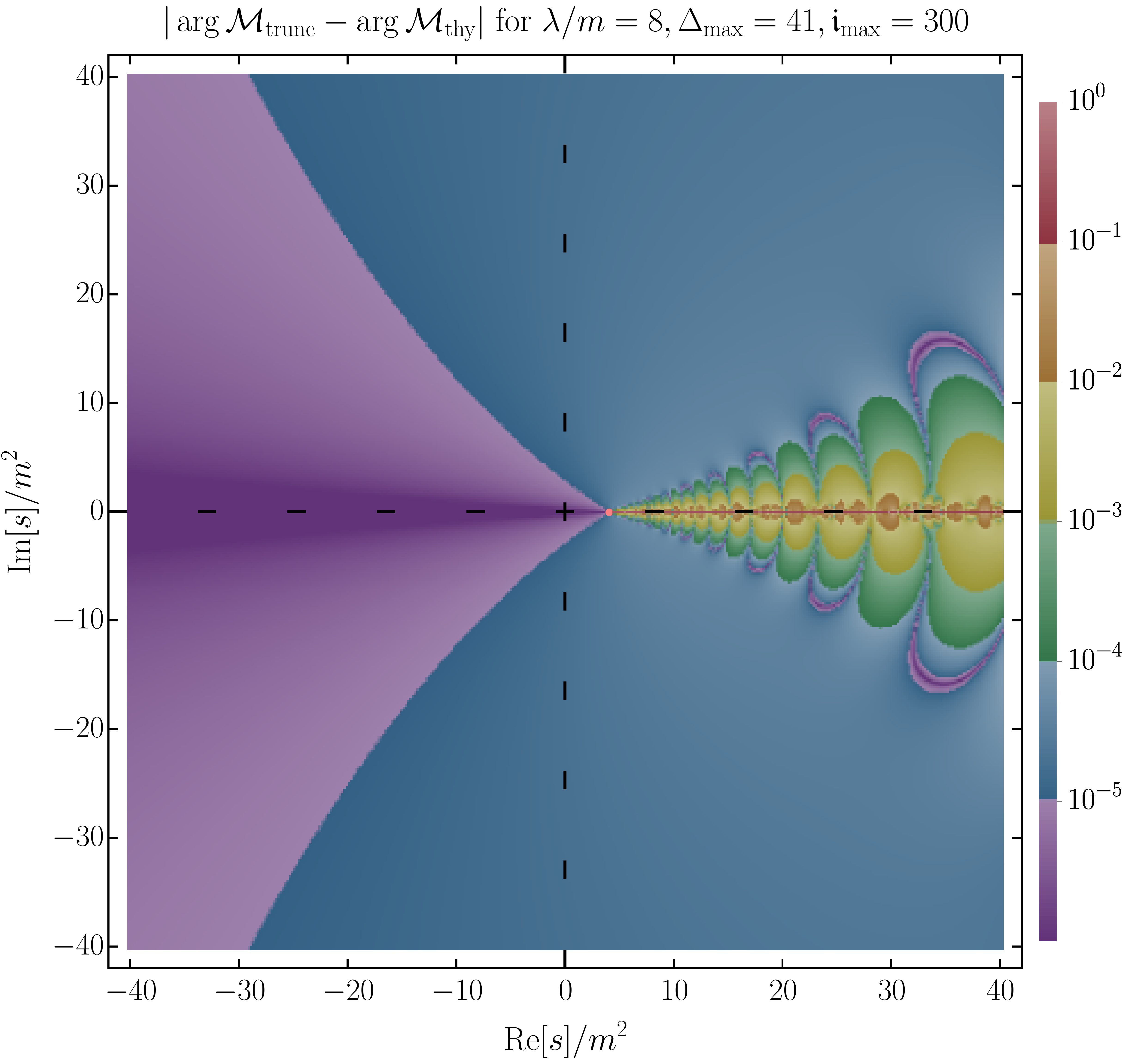}
    \caption{\footnotesize{\textbf{Left:} Comparison of the truncation and theoretical results for the absolute value of the $s$-flavor amplitude.  \textbf{Right:} The same comparison for the phase of the amplitude. Both of these plots are made with truncation parameters $\Dmax = 41$, $\imax = 300$ (6020 states). The convergence is fastest away from the physical region $s > 4 m^2$.  In the physical region the convergence can be improved by using smearing techniques such as those of Eq.~\eqref{eq:smeared_im_part} and Eq.~\eqref{eq:cutoff_re_part}, which produce the results show in Fig.~\ref{fig:sChanResiduals}.  To guide the eye there is a pink dot in each figure located at the threshold $s = 4 m^2$.}}
    \label{fig:complexSresiduals}
\end{figure}

So far we have only discussed truncation results for the $s$-flavor amplitude. However, Figs.~\ref{fig:sChanResiduals} and \ref{fig:complexSresiduals} are actually direct tests of the $t$- and $u$-flavor amplitudes, as well. This is most obvious for the $t$-flavor amplitude, as its truncation expression~\eqref{eq:tchannel_truncation} is \emph{identical} to that of the $s$-flavor expression~\eqref{eq:schannel_truncation} with $s \ra t$. For the $u$-flavor amplitude, in App.~\ref{app:ON_truncation_detail} we show explicitly that its truncation expression~\eqref{eq:uchannel_truncation} can be rewritten into a form identical to~\eqref{eq:schannel_truncation} with $s \ra u$.

In other words, for the $O(N)$ model at large $N$, the $t$- and $u$-flavor computations simply correspond to the $s$-flavor computation analytically continued to $s < 0$ (since physical amplitudes have $t,u <0$). From Figs.~\ref{fig:sChanResiduals} and \ref{fig:complexSresiduals}, we can therefore see that the convergence for these amplitudes is even better than that of the $s$-flavor case.

\section{Discussion} \label{sec:discussion}
This work provides a framework for numerically computing scattering amplitudes in a strongly-coupled quantum field theory.  Using lightcone conformal truncation, we can compute a collection of approximate energy eigenstates for a QFT; then employing the techniques outlined in this paper we may use these eigenstates to compute amplitudes.  The principal barrier in doing so is in finding a representation of the $S$-matrix that is as manifestly smooth as possible, so that when we discretize and insert approximate sums over eigenstates, we do not disrupt any delicate cancellation between zeroes and poles.  Physically, such a cancellation arises from the equations of motion, so it is unsurprising that the way to avoid this issue is to properly utilize the equations of motion, which we do via the Schwinger-Dyson equations.  We then arrive at Eq.~\eqref{eq:GeneralAmp}, which gives us access to the scattering amplitude for arbitrary external particle momenta.

One of the exciting aspects of this technique is that it allows us to numerically construct the amplitude for both real and complex external kinematics, which we demonstrate in Sec.~\ref{sec:truncation_results} by computing an amplitude on the complex $s$-plane.  This means that, in principle, our method can probe the complex analytic structure of the $S$-matrix.  However there is further work needed to fully understand this point. Indeed, in a continuum QFT branch points form at multi-particle thresholds; by passing through a cut one can encounter additional complex structure such as a resonance, as indicated by a pole off the real line. On the other hand, numerical methods---like those used in this paper---necessarily have a discrete spectrum and therefore, at finite truncation, produce a series of dense poles but not a fully fledged branch cut.\footnote{As an example, for a theory of real scalar fields, the truncation basis states may be normalized such that the Hamiltonian is fully real (i.e.\ with no complex phases).  This implies that all matrix elements in Eq.~\eqref{eq:GeneralAmp} will be real and thus that all complex analytic structure arises from the $(M_\alpha^2 - s - i \epsilon)^{-1}$ and $(M_\alpha^2 - t - i \epsilon)^{-1}$ propagators.  At finite truncation, the analytic structure will thus be a series of dense poles on the contours of real $s$ and $t$.} For this reason, it is unclear whether and how our truncation formula can access the full multisheeted structure of the amplitude.  In the concrete example of a resonance, we certainly expect that our method will reproduce the proper Breit-Wigner structure for real on-shell kinematics, but it is less obvious how well one can numerically reproduce a complex pole on a higher sheet (or its associated zero on the physical sheet).  It is conceivable, for example, that convergence is in fact better in the complex plane, and that this can be used to give a more precise measure of the mass and width of an unstable particle resonance.  These and similar questions represent important targets for understanding how to study and exploit complex analytic structure using Eq.~\eqref{eq:GeneralAmp}, but we leave them for future work.

In Figs.~\ref{fig:sChanResiduals} and \ref{fig:complexSresiduals}, we found that the convergence of our numerical results is noticeably faster for unphysical values of $s$, with $\lesssim 0.1\%$ error across most of the complex plane. This behavior is familiar from truncation studies of the analytic structure of correlation functions~\cite{Chen:2021bmm}, which were able to use the rapid convergence at unphysical energies to further improve truncation results in the physical regime, and it would be very interesting to generalize such an analysis to scattering amplitudes, as well. A more systematic understanding of the rate of convergence as a function of $\Dmax$ and $\imax$ would also allow these truncation results to be extrapolated to higher precision.

It is important to note that our method does not give access to the full complex $s$-$t$ plane. A related fact is that full \((s,t,u)\) crossing symmetry is not manifest in our formulation. Ultimately, both aspects originate from our choice to use LSZ to reduce only two of the four particles and work with the correlator \(\braket{\onmom{4}{}|T\{\phi(x_3)\phi(x_2)\}|\onmom{1}{}}\). This has the advantage that it allows us to directly use the one-particle states \(\bra{\onmom{4}{}}\) and \(\ket{\onmom{1}{}}\), which are easily singled out in truncation because they are isolated from the continuum spectrum. On the other hand, as \(\bra{\onmom{4}{}}\) and \(\ket{\onmom{1}{}}\) are physical states, this formulation necessarily demands that \(p_1\) and \(p_4\) be on-shell with positive energy, and thus that \(u\) be real and physical---see also Footnote~\ref{ftntan}. It is for this reason that our method takes the form of a fixed-\(u\) dispersion relation, see Eq.~\eqref{eq:dispersion_relation}. To access the full complex $s$-$t$ plane (with \(s+t+u= \sum_i m_i^2\), as always) it is sufficient to reduce out just one more particle.\footnote{The expression when reducing out three particles, after applying Schwinger-Dyson relations, reads 
\begin{align*}
    (\Box_3 & + m^2) (\Box_2 + m^2) (\Box_1 + m^2) \langle p_4 | T \phi_3 \phi_2 \phi_1 | 0 \rangle \\
    &= \langle p_4 | T J_3 J_2 J_1 | 0 \rangle -i \Big[\langle p_4 | T J_1 J_2' | 0 \rangle\delta^d (x_2 - x_3)+\langle p_4 | T J_2 J_3' | 0 \rangle\delta^d (x_3 - x_1)+\langle p_4 | T J_3 J_1' | 0 \rangle\delta^d (x_1 - x_2)\Big] \\
    &~~~~~ - \langle p_4 | J_3'' | 0 \rangle \de^d(x_1-x_3) \de^d(x_2-x_3) + \text{disconnected terms} .
\end{align*}
} If instead we reduce out all four particles we will have a manifestly crossing-symmetric expression, where the numerically stable correlator of interest takes the form
\begin{equation}
    (\Box_4+m^2)\cdots(\Box_1 + m^2)\braket{T\{\phi_4\phi_3\phi_2\phi_1\}} = \braket{T\{J_4J_3 J_2 J_1\}} + \text{contact terms}.
\end{equation}
It will be interesting in future work to explore if and how crossing symmetry can be utilized to e.g.\ improve convergence, as well as the relation to Mandelstam's conjectured double dispersion relation~\cite{Mandelstam:1958xc}.

In this work we have focused on $2 \to 2$ scattering in the large-$N$ $O(N)$ model in 2+1d.  This was in order to compare our truncation calculation against known results, but this same procedure can be used to compute the $S$-matrix in a wide variety of strongly-coupled, finite-$N$ theories. Obvious short-term targets include many systems in 1+1d~\cite{Lassig:1990xy,Katz:2013qua,Katz:2014uoa,Coser:2014lla,Rychkov:2014eea,Rychkov:2015vap,Bajnok:2015bgw,Anand:2017yij,Fitzpatrick:2019cif,Konik:2020gdi,Dempsey:2021xpf,Anand:2021qnd}, as well as 2+1d $\phi^4$ theory~\cite{Elias-Miro:2020qwz,Anand:2020qnp}, which have previously been studied with Hamiltonian truncation. There are also no conceptual hurdles (such as the sign and fermion-doubling problems that plague lattice realizations~\cite{Nielsen:1981hk,Troyer:2004ge}) to studying theories with fermions. Furthermore, one can potentially compute scattering amplitudes at finite density and temperature by applying LSZ to correlation functions in the background of high-energy eigenstates~\cite{Delacretaz:2022ojg}. In principle, moving to $3+1$ dimensions poses no conceptual challenges, though almost all deformations of free theories are marginal in that case, where it is currently unknown how to efficiently implement truncation methods.\footnote{Due to the logarithmic running, one anticipates slower convergence, and there remain open questions on the cancellation of UV divergences due to marginal operators.}
It is also worth emphasizing that while here we have used the specific method of LCT,  Eq.~\eqref{eq:GeneralAmp} can be implemented in any nonperturbative framework which computes energy eigenstates, such as the truncated conformal space approach~\cite{Yurov:1989yu,Yurov:1991my,Hogervorst:2014rta} or relativistic continuous matrix product states~\cite{Tilloy:2021hhb,Tilloy:2021yre}.

A crucial aspect of our prescription was the use of the equations of motion to relate the scattering amplitude to correlation functions involving a source $J$, such that the resulting expression~\eqref{eq:GeneralAmp} is manifestly smooth on-shell. In general, the LSZ prescription doesn't only hold for the field $\phi$, but for any operator $\Ocal$ with the same quantum numbers as the one-particle state $|\ve{p}\>$. In the case that the asymptotic state is a composite (bound) state, one generically expects the interpolating operator \(\mathcal{O}\) to be a composite operator built from the fundamental fields.\footnote{For example, in QCD the leading operators that interpolate the pions are quark bilinears $\sim \bar{q}\gamma_5q$.} In such a situation there are no obvious equations of motion which replace \((\Box + m^2)\mathcal{O} \to J_{\mathcal{O}}\), though one possibility would be to use the equations of motion for the consitutient fields.\footnote{A concrete setting in which to better understand this question is $\phi^4$ theory, where one can use an operator such as $\phi^3$ instead of $\phi$ in the original correlator.} In the case where the bound state arises as a (pseudo) Nambu-Goldstone boson, one could potentially also make use of Ward identities.

Note that in Eq.~\eqref{eq:eom_phi} we have explicitly defined the source $J$ in terms of the \emph{physical} mass~$m$ (i.e.~the eigenvalue of the first excited state of $H$). Generically, the physical mass differs from the \emph{bare} mass $m_0$ appearing in the action (which therefore typically appears in the equations of motion), in which case the source will contain a term proportional to this mass shift, $J \supset (m^2 - m_0^2) \phi$. This contribution to $J$ is crucial to cancel against corrections to the external propagators for particles 2 and 3 to ensure that Eq.~\eqref{eq:GeneralAmp} correctly reproduces the amplitude. In theories with UV divergences, this cancellation is more subtle and requires the inclusion of ``nonlocal'' or ``state-dependent'' counterterms~\cite{Elias-Miro:2017tup,Rutter:2018aog,Elias-Miro:2020qwz,Anand:2020qnp,Cohen:2021erm,EliasMiro:2021aof} in $V$ (and therefore $J$).

One asymptotic goal of any nonperturbative technique is to study strongly-coupled gauge theories, including quantum chromodynamics (QCD). There are two potential strategies for doing so. The obvious approach is to start with a free theory of quarks and gluons, then add gauge interactions as the deformation. However, there are important conceptual hurdles that must first be overcome: the interaction terms do not correspond to gauge-invariant local operators and are marginal in 3+1d. An alternative approach is to start with an interacting CFT (e.g.\ a Banks-Zaks fixed point) which contains the desired gauge theory as a subset of its degrees of freedom. The extra matter content can then be removed with large mass terms (which \emph{are} relevant, gauge-invariant local operators), leaving only the desired QFT at low energies. Of course the challenge with such an approach is the initial calculation of the CFT data, through the conformal bootstrap~\cite{Poland:2018epd} or other methods. Given the importance of this goal, we believe both approaches should be studied carefully in future work.

Our understanding of strongly-coupled QFT has advanced immensely over the past several decades, but we are far from having a complete and satisfying picture.  There are many simple physical phenomena for which we believe we have a descriptive theory, but we are unable to numerically compute the predictions of this theory and thus unable to make a faithful comparison to experiment.  In this work, we have tried to bridge one gap by outlining a new tool that may be used to compute real-time particle physics observables such as cross sections and decay rates. We hope this approach, and Hamiltonian methods more generally, will continue to be explored as an exciting avenue for connecting theory with experiment.

\begin{acknowledgments}
It is a pleasure to thank Olivier Delouche, Savas Dimopoulos, Liam Fitzpatrick, Max Hansen, Denis Karateev, Ami Katz, Markus Luty, Sasha Monin, Jo\~ao Penedones, and Riccardo Rattazzi for valuable discussions. J.O.T.\ would like to thank the University of Geneva, CERN, EPFL, and Perimeter Institute for hospitality during the completion of this work.
The work of B.H. is partially supported by the Swiss National Science Foundation under contract 200020-188671 and through the National Center of Competence in Research SwissMAP. 
The work of H.M.\ was supported by the Director, Office of Science, Office of High Energy Physics of the U.S. Department of Energy under the Contract No. DE-AC02-05CH11231, by the NSF grant PHY-1915314, by the JSPS Grant-in-Aid for Scientific Research JP20K03942, MEXT Grant-in-Aid for Transformative Research Areas (A) JP20H05850, JP20A203, and Hamamatsu Photonics, K.K. In addition H.M.\ is supported by the World Premier International Research Center Initiative (WPI) MEXT, Japan. This research was supported in part by Perimeter Institute for Theoretical Physics. Research at Perimeter Institute is supported by the Government of Canada through the Department of Innovation, Science and Economic Development and by the Province of Ontario through the Ministry of Research, Innovation and Science. The work of F.R.\ and M.W.\ is supported by the Swiss National Science Foundation under grants no. 200021-205016 and PP00P2-206149.

\end{acknowledgments}

\appendix

\section{\texorpdfstring{Truncation of the $O(N)$ model in more detail}{Truncation of the O(N) model in more detail}} \label{app:ON_truncation_detail}

In this appendix we go through the calculation of the $O(N)$ model scattering amplitude with lightcone conformal truncation in more detail so the interested reader may carefully track all the conventions and factors.  This parallels Ref.~\cite{Katz:2016hxp} closely, although there are minor differences in the notation and structure of the basis states.

\subsection{General conventions}

Our starting point is the Lagrangian for the $O(N)$ model in $d=2+1$:
\begin{equation}
    \Lcal = \Lcal^{\text{(CFT)}} + \Lcal^{(m)} + \Lcal^{(\lambda)} \equiv \half \del^\mu \phi^i \del_\mu \phi^i - \half m^2 \phi^i \phi^i - \frac{\lambda}{4 N} (\phi^i \phi^i) (\phi^j \phi^j).
\end{equation}
Note that all local operators in this work are taken to be normal-ordered. We use the lightcone coordinates
\be
x^\pm \equiv \fr{1}{\sqrt{2}}(x^0 \pm x^1), \quad x^\perp \equiv x^2,
\ee
with the metric $ds^2 = 2 dx^+ dx^- - dx^{\perp 2}$. Our theory is quantized on slices of fixed $x^+$, such that $x^-$ and $x^\perp$ are the ``spatial'' coordinates. To simplify the resulting expressions, we will use the shorthand notation:
\be
\dd \ve{x} \equiv dx^- \, dx^\perp, \quad \dd \ve{p} \equiv \fr{dp_- \, dp_\perp}{(2\pi)^2 2p_-}, \quad \de(\ve{p}-\ve{q}) \equiv 2p_- (2\pi)^2 \de(p_- - q_-) \de(p_\perp - q_\perp).
\ee
Note that all momentum integrals are restricted to the range $p_- \in [0,\infty)$, $p_\perp \in (-\infty,\infty)$.

Our first task is to construct the Hamiltonian $P_+$, which is the generator of translations in $x^+$,
\begin{equation}
    P_+ \equiv \int \dd \ve{x} \; T^+_{\; \; \; +} = \int \dd \ve{x} \; T_{-+},
\end{equation}
where $T_{\mu \nu}$ is the usual stress-energy tensor
\begin{equation}
    T_{\mu \nu} \equiv \del_\nu \phi^i \frac{\delta \Lcal}{\delta (\del^\mu \phi^i)} - g_{\mu \nu} \Lcal.
\end{equation}
The contributions to $P_+$ from the individual terms in the Lagrangian are therefore:
\be
\ba
    P_+^{\text{(CFT)}} &= \half \int \dd\ve{x} \, \del_\perp \phi^i \del_\perp \phi^i, \\
    \delta P_+^{(m)} &= \fr{m^2}{2} \int \dd\ve{x} \, \phi^i \phi^i, \\
    \delta P_+^{(\lambda)} &= \frac{\lambda}{4 N} \int \dd\ve{x} \, (\phi^i \phi^i)^2.
\ea
    \label{eq:Pplus_in_terms_of_phi}
\ee
Note that $\phi^i$ corresponds to the free massless field in the UV CFT, which has the mode expansion
\begin{equation} \label{eq:mode_expansion}
    \phi^i(x) = \int \dd\ve{p} \, \Big(e^{- i p \cdot x} a_{\ve{p}}^i + e^{i p \cdot x} a_{\ve{p}}^{\dagger i} \Big),
\end{equation}
where in the exponent $p_+$ is given by its massless on-shell value, $p_+ = \frac{p_\perp^2}{2 p_-}$. We have normalized the creation and annihilation operators such that they satisfy the commutation relation
\begin{equation}
    [a_{\ve{p}}^i, a_{\ve{q}}^{\dagger j}] = \de(\ve{p}- \ve{q}) \, \delta^{i j}.
\end{equation}
These modes create the single- and multi-particle states
\be
\ba
    | \ve{p}^i\> \equiv a_{\ve{p}}^{\dagger i} | 0 \>, \qquad | \ve{p}_1^{i_1} , \ldots , \ve{p}_n^{i_n} \> \equiv \frac{1}{\sqrt{n!}} a_{\ve{p}_1}^{\dagger i_1} \cdots a_{\ve{p}_n}^{\dagger i_n} | 0 \>,
\ea
\ee
which are thus normalized as
\be
\ba
    \< \ve{p}^i | \ve{q}^j \> &= \de(\ve{p} - \ve{q}) \, \delta^{i j}, \\
    \< \ve{p}_1^{i_1} , \ldots , \ve{p}_n^{i_n} | \ve{q}_1^{j_1} , \ldots , \ve{q}_n^{j_n} \> &= \frac{1}{n!} \sum_{\sigma \in S_n} \prod_{\ell = 1}^n \Big[ \delta( \ve{p}_\ell - \ve{q}_{\sigma_\ell} ) \, \delta^{i_\ell j_{\sigma_\ell}} \Big].
\ea
\ee

In the above, $| 0 \>$ is the free CFT vacuum and the multi-particle Fock space states are defined in the free theory. Because we are working in lightcone quantization, $|0\>$ remains the true vacuum even when we deform the theory by adding $\de P_+^{(m)}$ and $\de P_+^{(\lambda)}$ (see~\cite{Fitzpatrick:2018ttk} for further discussion). However, all excited states in the deformed theory will correspond to linear combinations of the multi-particle states, with no definite particle number.

Using the mode expansion we can rewrite the Hamiltonian contributions of Eq.~\eqref{eq:Pplus_in_terms_of_phi} in terms of creation and annihilation operators:
\be
\ba
    P_+^{\text{(CFT)}} &= \int \dd\ve{p} \, \frac{p_\perp^2}{2 p_-} a_{\ve{p}}^{\dagger i} a_{\ve{p}}^i, \\
    \delta P_+^{(m)} &= \int \dd\ve{p} \, \frac{m^2}{2 p_-} a_{\ve{p}}^{\dagger i} a_{\ve{p}}^i, \\
    \delta P_+^{(\lambda)} &= \frac{\lambda}{4 N} \int \dd\ve{p} \, \dd\ve{q} \, \dd\ve{k} \Bigg[ \frac{a_{\ve{p}}^{\dagger i} a_{\ve{q}}^{\dagger i} a_{\ve{k}}^j a_{\ve{p}+\ve{q}-\ve{k}}^j}{p_- + q_- - k_-} + \frac{2 a_{\ve{p}}^{\dagger i} a_{\ve{q}}^{\dagger j} a_{\ve{k}}^i a_{\ve{p}+\ve{q}-\ve{k}}^j}{p_- + q_- - k_-} \\
    & \hspace{3.5cm} + \frac{2 a_{\ve{p}}^{\dagger i} a_{\ve{q}}^{\dagger i} a_{\ve{k}}^{\dagger j} a_{\ve{p}+\ve{q}+\ve{k}}^j}{p_- + q_- + k_-}
    + \frac{2 a_{\ve{p}+\ve{q}-\ve{k}}^{\dagger i} a_{\ve{p}}^{i} a_{\ve{q}}^{j} a_{\ve{k}}^j}{p_- + q_- + k_-}
    \Bigg].
\ea
\label{eq:Pplus_in_terms_of_a}
\ee
Note the absence of any terms consisting solely of $a$'s or solely of $a^\dag$'s. This is due to the mandatory positivity of $p_-$ for all physical particles in lightcone quantization, which kills all matrix elements involving creation of particles from nothing.  We thus see explicitly that the deformations to the Hamiltonian will not mix the vacuum $| 0 \>$ with any other state.

\subsection{Choice of CFT basis states} \label{app:two_particle_states}

In general, to implement LCT we need to construct all primary operators in the UV CFT of free field theory with $\De \leq \Dmax$. For the results of this paper, however, we will only need flavor-singlet two-particle operators of the schematic form
\begin{equation} \label{eq:operator_in_terms_of_phi}
    \Ocal(x) \sim \fr{1}{\sqrt{N}} \del_-^\# \del_\perp^\# \phi^i(x) \del_-^\# \del_\perp^\# \phi^i(x),
\end{equation}
where these operators are normalized such that their two-point functions are finite in the limit $N \to \infty$. Following Eq.~\eqref{eq:truncation_states}, we can write the states created by these operators as:
\begin{equation} \label{eq:truncation_states_3d_pos_space}
    | \Ocal; \ifrak ; \ve{p} \> \equiv \int d\mu^2 g_\ifrak(\mu) \int d^3x \, e^{- i p \cdot x} \Ocal(x) | 0 \>,
\end{equation}
where in the exponent $p_+ = \fr{\mu^2 + p_\perp^2}{2p_-}$.

Substituting the mode expansion~\eqref{eq:mode_expansion} into~\eqref{eq:truncation_states_3d_pos_space}, we can rewrite these two-particle basis states in terms of Fock space states:
\be \label{eq:truncation_states_3d_mom_space}
\ba
    | \Ocal; \ifrak; \ve{p} \> \equiv \int\! d\mu^2 & g_\ifrak(\mu) \int \! \dd\ve{p}_1 \, \dd\ve{p}_2 (2 \pi) \delta \Big( \mu^2 - (p_1 + p_2)^2 \Big) \de(\ve{p}-\ve{p}_1-\ve{p}_2) \widetilde{F}_\De(\ve{p}_1,\ve{p}_2) | \ve{p}_1^i , \ve{p}_2^i \>,
\ea
\ee
where $\widetilde{F}_\De(\ve{p}_1,\ve{p}_2)$ is a polynomial in the particle momenta. In a flavor-singlet state, the two particles are indistinguishable so the polynomials $\widetilde{F}_\De$ must be symmetric (i.e.\ they must be invariant under $\ve{p}_1 \leftrightarrow \ve{p}_2$).

In the $O(N)$ model, all flavor-singlet two-particle primary operators correspond to conserved currents (one for each even spin), each with only two independent degrees of freedom.  These degrees of freedom can be organized into the two parity eigenstates under $p_{i\perp} \mapsto -p_{i\perp}$, and the $2 \to 2$ scattering amplitudes we compute in this work only require the parity-even states. As a result, the polynomials $\widetilde{F}_\De$ are uniquely labeled by the scaling dimension $\De$ of the associated operator $\Ocal$ and can be written as functions only of $p_{i-}$ (i.e.\ not of $p_{i\perp}$).  Finally, as discussed in~\cite{Katz:2016hxp}, the mass deformation $\de P_+^{(m)}$ lifts one linear combination of primary operators out of the low-energy spectrum, so it is simpler in practice to modify our basis to the set of polynomials $\widetilde{F}_\De$ that are orthogonal to this lifted state.

Details of all of these issues are presented in~\cite{Katz:2016hxp}, but the main takeaway here is that we have one CFT basis state for each odd scaling dimension $\De \geq 3$.  The relevant polynomials for these states are given in Eq.~(E.19) of~\cite{Katz:2016hxp} which we reproduce here:
\be
\widetilde{F}_\De(\ve{p}_1,\ve{p}_2) \equiv \fr{1}{\sqrt{N}} \frac{\sqrt{2 (\De-1) \Gamma(\De - 2) \Gamma (\De + 1)}}{\Gamma(\De - \half)} p_{1-} p_{2-} (p_{1-} + p_{2-})^{\De-3} P^{( \frac{3}{2}, \frac{3}{2})}_{\De - 3}\bigg(\frac{p_{2-} - p_{1-}}{p_{1-} + p_{2-}}\bigg),
\ee
for $\De = 3,5,7,\ldots,$ where $P^{(a, b)}_{c}(x)$ is a Jacobi polynomial. We truncate this basis by only keeping states with $\De \leq \Dmax$.

We also must choose smearing functions $g_\ifrak(\mu)$, and here we use a different convention than in~\cite{Katz:2016hxp}.  For ease of computation, we choose top-hat window functions of the form:
\begin{equation}
    g_\ifrak(\mu) =
    \begin{cases}
    \frac{1}{\sqrt{2(\mu_{\ifrak+1} - \mu_\ifrak)}} & \mu_\ifrak \leq \mu < \mu_{\ifrak + 1},  \\
    0 & \text{otherwise}.
    \end{cases}
\end{equation}
Because we are focused on resolving the scattering amplitude in the IR, we choose to use narrower windows (i.e.\ smaller $\mu_{\ifrak + 1} - \mu_\ifrak$) at smaller values of $\mu$:
\be
\mu_\ifrak \equiv \LambdaUV \exp\Big[{\fr{\ifrak-1-\imax}{\boldsymbol{\de}\ifrak_{\textrm{ref}}}}\Big],
\ee
where $\LambdaUV$ is the resulting UV cutoff and $\boldsymbol{\de}\ifrak_{\textrm{ref}}$ is a free parameter controlling the relative size of adjacent windows (in this work we use $\LambdaUV = 200m$ and $\boldsymbol{\de}\ifrak_{\textrm{ref}}=25$).
To truncate, we choose some $\ifrak_{\rmmax}$ to specify the number of windows and keep only the finite number of states with $\ifrak \leq \ifrak_{\rmmax}$.

With these choices of $\widetilde{F}_\De$ and $g_\ifrak$, we can now compute the inner product for our basis states,
\begin{equation}
    \< \Ocal; \ifrak; \ve{p} | \Ocal'; \jfrak; \ve{p}' \> = \delta_{\Ocal \Ocal'} \delta_{\frakI \frakJ} \de(\ve{p} - \ve{p}').
\end{equation}
The approximate resolution of the identity in the truncated two-particle sector is therefore:
\begin{equation}
    \mathbb{I}_2 \equiv \sum_{\De=3}^{\Dmax} \sum_{\ifrak = 0}^{\ifrak_{\rmmax}} \int \dd\ve{p} \, | \Ocal; \ifrak; \ve{p} \> \< \Ocal ; \ifrak; \ve{p} |.
\end{equation}

\subsubsection*{Higher particle number states}

For the computation of the various flavor amplitudes, we will also need three- and four-particle basis states. In general, we could just construct all primary operators with the appropriate particle number and $O(N)$ representation, but as we will see, at large $N$ only a two-particle subset of each state is affected by the interaction. We therefore only need to consider basis states which factorize into the following form:
\be
|\Ocal;\ifrak;\ve{p}-\ve{q}\> \otimes |\ve{q}^i\>, \qquad |\Ocal;\ifrak;\ve{p}-\ve{q}_1-\ve{q}_2\> \otimes |\ve{q}_1^i,\ve{q}_2^j\>,
\ee
i.e., two-particle CFT basis states plus one or more ``spectator'' particles. We have chosen this basis purely for convenience in this particular model, in order to focus on how one extracts the $S$-matrix from truncation data, rather than the details of how particular intermediate states are reproduced by CFT operators.

\subsection{Matrix elements for truncation} \label{app:matrix_elements}

We now need to construct the invariant mass-squared matrix $\mathbb{M}^2 \equiv 2 P_+ P_- - P_\perp^2$ in our truncated basis. The spatial momentum generators $P_-$ and $P_\perp$ are unaffected by the deformations to the UV CFT, so the only corrections come from $P_+$. The matrix elements of $P_+$ for flavor-singlet two-particle basis states can be directly computed from the Fock space mode expressions for $P_+$ in Eq.~\eqref{eq:Pplus_in_terms_of_a}, and the results are (to leading order in $1/N$) \cite{Katz:2016hxp}:
\be
\ba
    \< \Ocal; \ifrak; \ve{p} | \mathbb{M}^2_{\mathrm{(CFT)}} | \Ocal' ; \jfrak; \ve{p}' \> &= \de_{\Ocal\Ocal'} \de_{\ifrak\jfrak} \de(\ve{p} - \ve{p}') \cdot \frac{1}{3} \frac{\mu_{\frakI+1}^3 - \mu_\frakI^3}{ \mu_{\frakI+1} - \mu_\frakI}, \\
    \< \Ocal; \ifrak; \ve{p} | \de\mathbb{M}^2_{(m)} | \Ocal' ; \jfrak; \ve{p}' \> &= \de_{\ifrak\jfrak} \de(\ve{p}-\ve{p}') \cdot \frac{8m^2}{3} (\De_- - 1) \sqrt{\frac{\De_-(\De_- - 2)}{\De_+(\De_+-2)}}, \\
    \< \Ocal; \ifrak; \ve{p} | \de\mathbb{M}^2_{(\lambda)} | \Ocal' ; \jfrak; \ve{p}' \> &= \de(\ve{p}-\ve{p}') \cdot \frac{\lambda}{2N} \< \Ocal; \ifrak; \ve{p} | \vec{\phi}^{\,2}(0) \> \< \vec{\phi}^{\,2}(0) | \Ocal' ; \jfrak; \ve{p}' \>,
\ea
\label{eq:comp_trunc_mat_elems}
\ee
where $\De_- = \rmmin (\De, \De')$ and $\De_+ = \rmmax (\De, \De')$.  Here we have written the matrix elements for the quartic interaction in a way that makes manifest the fact that it is a projector onto a single state: $| \vec{\phi}^{\,2}(0) \>$.  The overlaps of our basis with this state are given by \cite{Katz:2016hxp}:
\begin{equation} \label{eq:overlap_CFT_basis_phiSq}
    \frac{1}{\sqrt{N}} \< \vec{\phi}^{\,2}(0) | \Ocal; \ifrak; \ve{p} \> = \sqrt{\fr{\mu_{\ifrak + 1} - \mu_\ifrak}{\pi \De(\De-2)}}.
\end{equation}

Once we have constructed the resulting truncated $\mathbb{M}^2$ for a given choice of $\Dmax$ and $\imax$, we can diagonalize that finite-dimensional matrix to obtain the flavor-singlet two-particle approximate energy eigenstates,
\be
|M_\alpha^2; \ve{p}\> = \sum_{\substack{\De \leq \Dmax \\ \ifrak \, \leq \, \imax}} C^{\Ocal;\ifrak}_\alpha |\Ocal; \ifrak; \ve{p} \>.
\ee

For states with higher particle number, the analogous matrix elements for $\mathbb{M}^2$ are diagonal with respect to the spectator particles at leading order in $1/N$,
\be
\ba
    \< \ve{p}_n^{i_n},\ldots,\ve{p}_1^{i_1} | \mathbb{M}^2_{\mathrm{(CFT)}} | \ve{p}_1^{\prime j_1},\ldots,\ve{p}_n^{\prime j_n} \> &= \< \ve{p}_n^{i_n},\ldots,\ve{p}_1^{i_1} | \ve{p}_1^{\prime j_1},\ldots,\ve{p}_n^{\prime j_n} \> \cdot 2p_-\sum_{i=1}^n \fr{p_{i\perp}^2}{2p_{i-}}, \\
    \< \ve{p}_n^{i_n},\ldots,\ve{p}_1^{i_1} | \de\mathbb{M}^2_{(m)} | \ve{p}_1^{\prime j_1},\ldots,\ve{p}_n^{\prime j_n} \> &= \< \ve{p}_n^{i_n},\ldots,\ve{p}_1^{i_1} | \ve{p}_1^{\prime j_1},\ldots,\ve{p}_n^{\prime j_n} \> \cdot 2p_- \sum_{i=1}^n \fr{m^2}{2p_{i-}}, \\
    \< \ve{p}_n^{i_n},\ldots,\ve{p}_1^{i_1} | \de\mathbb{M}^2_{(\lambda)} | \ve{p}_1^{\prime j_1},\ldots,\ve{p}_n^{\prime j_n} \> &= \order(1/N).
\ea
\ee
The resulting multi-particle energy eigenstates therefore factorize into the two-particle eigenstates plus spectators,
\be
|M_\alpha^2; \ve{p} - {\textstyle\sum\nolimits_{i}\ve{p}_i}\> \otimes |\ve{p}_1^{i_1},\ldots,\ve{p}_n^{i_n}\>_{\textrm{free}},
\ee
where we have added the subscript to indicate that these spectators correspond to free Fock space states.

For the $u$-flavor amplitude, we will also need to explicitly compute the $\order(1/\sqrt{N})$ matrix element between the one-particle state and a three-particle state,
\be
\<\ve{q}^i| \otimes \<\Ocal;\ifrak;\ve{p}-\ve{q}| \de\mathbb{M}^2_{(\lambda)} | \ve{p}^{\prime j} \> = \de^{ij}\de(\ve{p}-\ve{p}') \cdot \fr{\lambda}{N} \<\Ocal;\ifrak;\ve{p}-\ve{q}| \vec{\phi}^{\,2}(0) \>.
\label{eq:MatrixElement1to3}
\ee

\subsection{Calculation of amplitude}

Once we have diagonalized the truncated Hamiltonian and obtained the associated energy eigenstates, we can compute the scattering amplitude via Eq.~\eqref{eq:amp_arbitrary_flavor}, which we reproduce here for convenience:
\be\label{eq:amp_arbitrary_flavor_app}
\ba
    \Mcal^{i j k \ell}(s,t) &= \frac{1}{Z} \Bigg[ \sum_{\alpha} \bigg( \frac{1}{M_\alpha^2 - s - i \epsilon} \< \onmom{4}{\ell} | J^k | M_\alpha^2 ; \ve{p}_1 + \ve{p}_2 \> \< M_\alpha^2 ; \ve{p}_1 + \ve{p}_2 | J^j | \onmom{1}{i} \> \\
    & \quad + \frac{1}{M_\alpha^2 - t - i \epsilon} \< \onmom{4}{\ell} | J^j | M_\alpha^2 ; \ve{p}_1 - \ve{p}_3 \> \< M_\alpha^2 ; \ve{p}_1 - \ve{p}_3 | J^k | \onmom{1}{i} \> \bigg) + \< \onmom{4}{\ell} | J'^{\,jk} | \onmom{1}{i} \> \Bigg],
\ea
\ee
where $J^i = -\fr{\lambda}{N} \vec{\phi}^{\,2} \phi^i$ and $J'^{\, ij} = -\fr{\lambda}{N}(2\phi^i \phi^j + \de^{ij} \vec{\phi}^{\,2})$. Here we proceed through this calculation in more detail for all three flavor amplitudes. While the precise details of this calculation are specific to the $O(N)$ model in the large $N$ limit, this example is intended to clarify the more general overall structure of how truncation reproduces the $S$-matrix via Eq.~\eqref{eq:amp_arbitrary_flavor_app}.

\subsubsection*{\texorpdfstring{$s$}{s}-flavor amplitude}

First, let's compute the $s$-flavor amplitude to leading order in $1/N$, which we obtain by contracting the flavor indices of particles 1 and 2 and particles 3 and 4,
\be
\fr{1}{N} \Mcal^{iijj}(s,t) \simeq \Mcal(s),
\ee
where we use $\simeq$ to indicate ``equivalent up to terms subleading in $1/N$.''

At large $N$, there are a few immediate simplifications in this calculation. First, as mentioned above, particle-number-changing matrix elements are suppressed in this limit, such that the single-particle state is unchanged by the quartic interaction at leading order in $N$. We therefore have
\be
|\ve{p}^i\> \simeq |\ve{p}^i\>_{\textrm{free}}, \quad Z \simeq 1,
\ee
such that we can easily evaluate the $J'$ term in Eq.~\eqref{eq:amp_arbitrary_flavor_app} to obtain
\be
\fr{1}{N} \< \onmom{4}{j} | J'^{\,ij} | \onmom{1}{i} \> \simeq -2\lambda,
\ee
which exactly reproduces the $\order(\lambda)$ term in $\Mcal(s)$ from Eq.~\eqref{eq:amp_single_flavor}.

\begin{figure}
    \centering
    \includegraphics[width=0.75\linewidth]{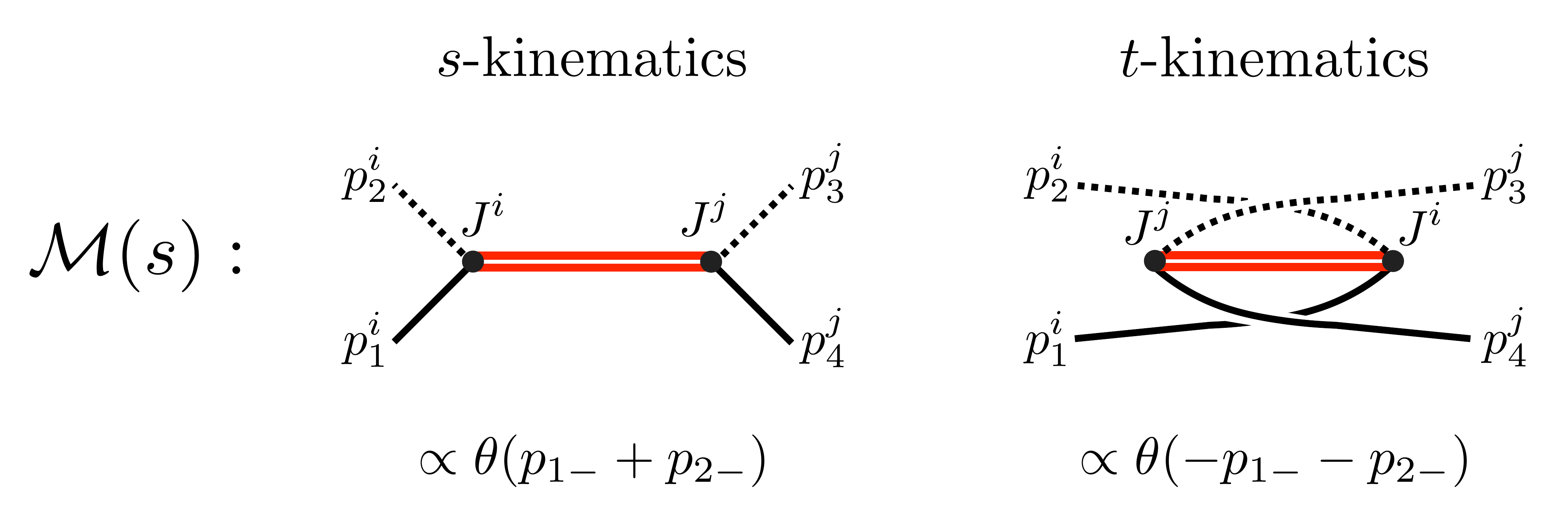}
    \caption{The $s$-flavor scattering amplitude receives contributions from two sets of intermediate states in Eq.~\eqref{eq:amp_arbitrary_flavor_app}: two-particle flavor-singlet energy eigenstates for the $s$-kinematics term (left) and four-particle states containing two spectators for the $t$-kinematics term (right).}
    \label{fig:schannel_amp_detailed}
\end{figure}

Turning to the $JJ$ terms in~\eqref{eq:amp_arbitrary_flavor_app}, the next simplification at large $N$ is that the $s$- and $t$-kinematics terms each receive a contribution from a single sector: flavor-singlet two-particle energy eigenstates for the $s$-kinematics term and four-particle states composed of two-particle eigenstates and two spectators for the $t$-kinematics term. These two contributions are shown schematically in Fig.~\ref{fig:schannel_amp_detailed}.

Let's go through these two contributions more carefully to understand their physical significance. The $s$-kinematics term corresponds to the ordering $\<\ve{p}_4|J(x_3) J(x_2)|\ve{p}_1\>$ in our LSZ formula~\eqref{eq:LSZinTermsOfJJ} (i.e.~$J(x_2)$ acts before $J(x_3)$). As a result, the two incoming particles 1 and 2 are both annihilated and two intermediate particles are created, which reproduce the chain of loops in the left diagram of Fig.~\ref{fig:ON_amp_diagram}. We thus have the $s$-kinematics contribution
\be
    \Mcal(s) \supset \fr{1}{N} \sum_{\alpha\in\{n=2\}} \frac{\< \onmom{4}{j} | J^j(0) | M_\alpha^2 ; \ve{p}_1 + \ve{p}_2 \> \< M_\alpha^2 ; \ve{p}_1 + \ve{p}_2 | J^i(0) | \onmom{1}{i} \>}{M_\alpha^2 - s - i \epsilon} .
    \label{eq:SChannel_2Part}
\ee
To evaluate this contribution, we need to compute the $J$ matrix element
\be
\ba
\< M_\alpha^2 ; \ve{q} | J^i(0) | \ve{p}^i \> &= -\fr{\lambda}{N} \< M_\alpha^2 ; \ve{q} | \vec{\phi}^{\,2} \phi^i(0) | \ve{p}^i \> \\
&\simeq -\lambda \< M_\alpha^2 ; \ve{q} | \vec{\phi}^{\,2}(0) \>,
\ea
\ee
which can be obtained from Eq.~\eqref{eq:overlap_CFT_basis_phiSq}.

Note that the two-particle intermediate states in~\eqref{eq:SChannel_2Part} are on-shell physical states, which means they must have physical energies. In particular, these states must have $p_{1-} + p_{2-} > 0$. This contribution to the amplitude is therefore proportional to $\theta(p_{1-} + p_{2-})$, as shown in Fig.~\ref{fig:schannel_amp_detailed}.

The $t$-kinematics term instead corresponds to the ordering $\<\ve{p}_4|J(x_2) J(x_3)|\ve{p}_1\>$. The associated intermediate states contain four particles: a two-particle energy eigenstate (which will reproduce the chain of loops) and two spectators (which will correspond to particles 1 and 4). We can therefore rewrite the resulting four-particle resolution of the identity in~\eqref{eq:amp_arbitrary_flavor_app} as
\be
\ba
\sum_{\alpha\in\{n=4\}} \!\!\!\! | M_\alpha^2 ; \ve{p}_1 \!- \!\ve{p}_3 \>\< M_\alpha^2 ; \ve{p}_1\! - \!\ve{p}_3| & \ra \!\!\!\! \sum_{\alpha\in\{n=2\}} \int \dd\ve{q} \, \dd\ve{q}_1 \, \dd\ve{q}_2 \, \de\big((\ve{p}_1 - \ve{p}_3) - (\ve{q} + \ve{q}_1 + \ve{q}_2)\big) \\
& \hspace{1.7cm} \times \!\big(| M_\alpha^2 ; \ve{q} \> \!\otimes\! |\ve{q}_1^k,\ve{q}_2^\ell\>_{\textrm{free}} \big) \big( {}_{\textrm{free}} \<\ve{q}_1^k,\ve{q}_2^\ell| \!\otimes\! \< M_\alpha^2 ; \ve{q} |  \big).
\ea
\ee
We need to evaluate the following $J$ matrix element involving these intermediate states,
\be
\ba
{}_{\textrm{free}} \<\ve{q}_1^k,\ve{q}_2^\ell|\otimes\< M_\alpha^2 ; \ve{q} | J^j(0) | \ve{p}^i \> &= -\fr{\lambda}{N} \, {}_{\textrm{free}} \<\ve{q}_1^k,\ve{q}_2^\ell|\otimes\< M_\alpha^2 ; \ve{q} | \vec{\phi}^{\,2}\phi^j(0) | \ve{p}^i \> \\
&\simeq -\fr{\lambda}{\sqrt{2}N} \< M_\alpha^2 ; \ve{q} | \vec{\phi}^{\,2}(0) \> \Big[ \de^{ik} \de^{j\ell} \de(\ve{p}-\ve{q}_1) + \de^{i\ell} \de^{jk} \de(\ve{p}-\ve{q}_2) \Big].
\ea
\ee
Given these matrix elements, the resulting numerator of the $t$-kinematics term can be written as
\be
\ba
&\big(\< \onmom{4}{j} | J^i(0) | M_\alpha^2 ; \ve{q} \> \otimes |\ve{q}_1^k,\!\ve{q}_2^\ell\>_{\textrm{free}} \big) \! \big( {}_{\textrm{free}} \<\ve{q}_1^k,\!\ve{q}_2^\ell|\otimes\< M_\alpha^2 ;\ve{q} | J^j(0) | \onmom{1}{i} \>\big) \\
& \qquad \simeq \fr{\lambda^2}{2} \< \vec{\phi}^{\,2}(0)|M_\alpha^2; \ve{q}\>\< M_\alpha^2 ; \ve{q} | \vec{\phi}^{\,2}(0) \> \Big[ \de(\ve{p}_1-\ve{q}_1) \de(\ve{p}_4-\ve{q}_2) + \de(\ve{p}_1-\ve{q}_2) \de(\ve{p}_4-\ve{q}_1) \Big].
\ea
\ee
We thus obtain the $t$-kinematics contribution
\be
\ba
\Mcal(s) &\supset \fr{\lambda^2}{N} \sum_{\alpha\in\{n=2\}} \int \dd\ve{q} \, \de\big((\ve{p}_1 - \ve{p}_3) - (\ve{q} + \ve{p}_1 + \ve{p}_4)\big) \frac{\< \vec{\phi}^{\,2}(0)|M_\alpha^2; \ve{q}\>\< M_\alpha^2 ; \ve{q} | \vec{\phi}^{\,2}(0) \>}{M_{(n=4)}^2 - t - i \epsilon} \\
&= \fr{\lambda^2}{N} \left( \fr{p_{1-} - p_{3-}}{-p_{1-}-p_{2-}} \right) \sum_{\alpha\in\{n=2\}} \frac{\< \vec{\phi}^{\,2}(0)|M_\alpha^2; -\ve{p}_1-\ve{p}_2\>\< M_\alpha^2 ; -\ve{p}_1-\ve{p}_2 | \vec{\phi}^{\,2}(0) \>}{M_{(n=4)}^2 - t - i \epsilon}
\ea
\ee
where $M_{(n=4)}^2$ refers to the invariant mass of the full four-particle state (singlet + spectators),
\be
\ba
M_{(n=4)}^2 &\equiv \big(\tfrac{M_\alpha^2 + (p_{1\perp} + p_{2\perp})^2}{-p_{1-}-p_{2-}} + \tfrac{p_{1\perp}^2 + m^2}{p_{1-}} + \tfrac{p_{4\perp}^2 + m^2}{p_{4-}}\big)(p_{1-}-p_{3-}) - (p_{1\perp}-p_{3\perp})^2 \\
&=\left(\fr{p_{1-}-p_{3-}}{-p_{1-}-p_{2-}}\right)(M_\alpha^2 - s) + t.
\ea
\ee
We can therefore rewrite the $t$-kinematics contribution in the more intuitive form
\be\label{eq:SChannel_4Part}
\Mcal(s) \supset \fr{\lambda^2}{N} \sum_{\alpha\in\{n=2\}} \frac{\< \vec{\phi}^{\,2}(0)|M_\alpha^2; -\ve{p}_1-\ve{p}_2\>\< M_\alpha^2 ; -\ve{p}_1-\ve{p}_2 | \vec{\phi}^{\,2}(0) \>}{M_\alpha^2 - s - i \epsilon}.
\ee
This exactly matches the $s$-kinematics contribution, except with $p_i \ra - p_i$, and therefore is instead proportional to $\theta(-p_{1-}-p_{2-})$, as indicated in Fig.~\ref{fig:schannel_amp_detailed}.

Naively, this $t$-kinematics contribution might seem somewhat unphysical. However, while particles 1 and 4 correspond to physical external states and therefore must have physical momenta (i.e.~$p_{1,4-} > 0$), particles 2 and 3 simply correspond to the Fourier transform of insertions of the source $J$, such that there is no equivalent restriction on $p_{2,3-}$. In particular, nowhere in Eq.~\eqref{eq:amp_arbitrary_flavor_app} is it specified that particle 2 is incoming and particle 3 is outgoing, this is instead indicated by the sign of their energies. We therefore need \emph{both} the $s$- and $t$-kinematics terms to ensure the full analytic structure in $p_2$ and $p_3$ of our partially-reduced LSZ reduction formula~\eqref{eq:LSZinTermsOfJJ}, giving us the final expression
\be\label{eq:SChannel_Full}
    \Mcal(s) \simeq \fr{\lambda^2}{N} \sum_{\alpha\in\{n=2\}} \frac{\big\< \vec{\phi}^{\,2}(0) \big| M_\alpha^2 ; |\ve{p}_1 + \ve{p}_2| \, \big\> \big\< M_\alpha^2 ; |\ve{p}_1 + \ve{p}_2| \, \big| \vec{\phi}^{\,2}(0) \big\>}{M_\alpha^2 - s - i \epsilon} - 2\lambda.
\ee
Eq.~\eqref{eq:SChannel_Full} explicitly reproduces the formulation of the $s$-flavor amplitude in terms of the spectral density of $\vec{\phi}^{\,2}$,
\be
\Mcal(s) = \fr{\lambda^2}{N} \int ds' \fr{\rho_{\vec{\phi}^{\,2}}(s')}{s'-s-i\epsilon} - 2\lambda,
\ee
which is simply a particular instance of the general dispersion relation~\eqref{eq:dispersion_relation}.

In practice, to compute the $s$-flavor amplitude via truncation we can choose to keep all external momenta physical, such that $p_{1,2-}>0$ and we only need the $s$-kinematics term and $J'$ term in~\eqref{eq:amp_arbitrary_flavor_app}. This is what we have done to obtain the results presented in Sec.~\ref{sec:truncation_results}.

\subsubsection*{\texorpdfstring{$t$}{t}-flavor amplitude}

Because Eq.~\eqref{eq:amp_arbitrary_flavor_app} is manifestly crossing symmetric under $p_2 \lra -p_3$, $J^j \lra J^k$, the calculation of the $t$-flavor amplitude
\be
\fr{1}{N} \Mcal^{ijij}(s,t) \simeq \Mcal(t),
\ee
is virtually identical to that of the $s$-flavor case, albeit with one conceptual difference we discuss below.

In particular, the $JJ$ terms receive the same contributions as above but with $s \lra t$: four-particle states for the $s$-kinematics term and two-particle states for the $t$-kinematics term, as shown schematically in Fig.~\ref{fig:tchannel_amp_detailed}. The requirement that the intermediate states are physical again imposes a restriction on the external particles' lightcone momenta: $p_{1-} - p_{3-} < 0$ for the $s$-kinematics term and $p_{1-} - p_{3-} > 0$ for the $t$-kinematics term.

\begin{figure}
    \centering
    \includegraphics[width=0.75\linewidth]{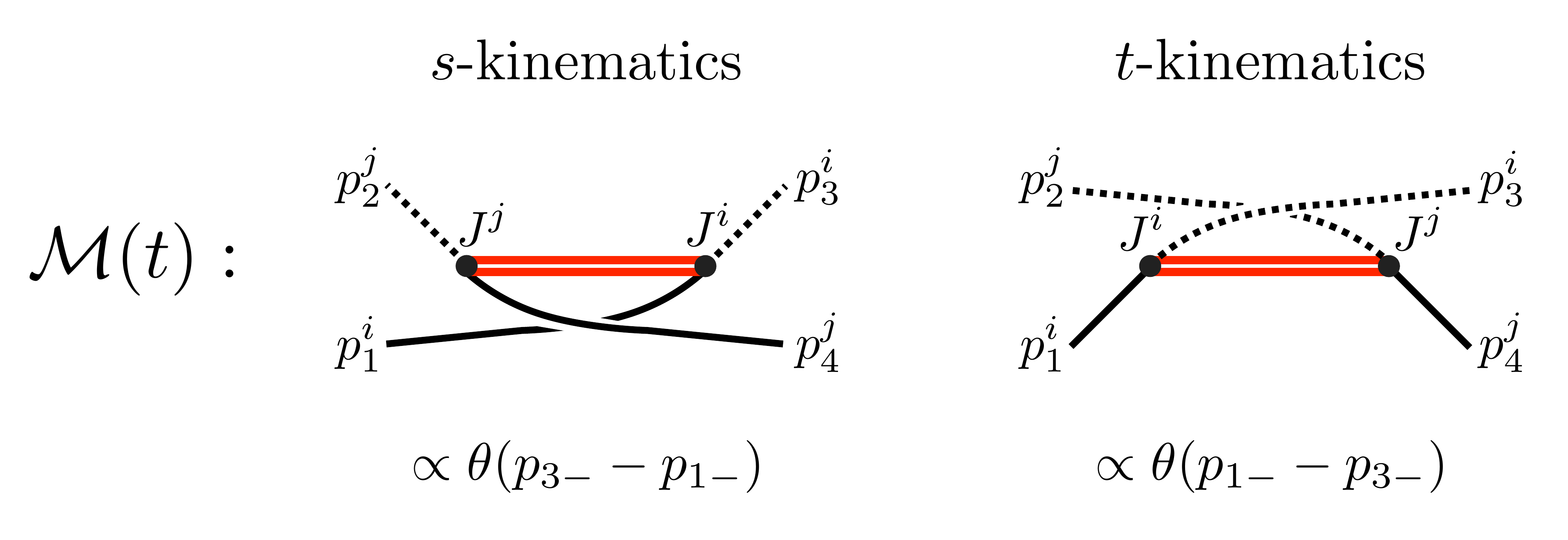}
    \caption{The $t$-flavor scattering amplitude receives contributions from two sets of intermediate states in Eq.~\eqref{eq:amp_arbitrary_flavor_app}: four-particle states containing two spectators for the $s$-kinematics term (left) and two-particle flavor-singlet energy eigenstates for the $t$-kinematics term (left).}
    \label{fig:tchannel_amp_detailed}
\end{figure}

Following the same manipulations as for the $s$-flavor amplitude, we obtain the final expression
\be\label{eq:TChannel_Full}
    \Mcal(t) \simeq \fr{\lambda^2}{N} \sum_{\alpha\in\{n=2\}} \frac{\big\< \vec{\phi}^{\,2}(0) \big| M_\alpha^2 ; |\ve{p}_1 - \ve{p}_3| \, \big\> \big\< M_\alpha^2 ; |\ve{p}_1 - \ve{p}_3| \, \big| \vec{\phi}^{\,2}(0) \big\>}{M_\alpha^2 - t - i \epsilon} - 2\lambda.
\ee
Analogously to the $s$-flavor case, in practice we can choose to work in a frame such that $p_{1-} > p_{3-}$, in which case we only need the $t$-kinematics term and the $J'$ term to compute the $t$-flavor amplitude via truncation.

While the contributions to Eq.~\eqref{eq:TChannel_Full} are structurally the same as in~\eqref{eq:SChannel_Full}, there is one important distinction in practice. For physical external momenta we have $s \geq 4m^2$ and therefore encounter the explicit poles $s=M_\alpha^2$ due to intermediate states when computing the amplitude. As discussed in Sec.~\ref{sec:truncation_results}, we therefore need to smear $\Mcal(s)$ in order to obtain a (relatively) smooth function of $s$. However, $t \leq 0$ for physical momenta, thus Eq.~\eqref{eq:TChannel_Full} never encounters any explicit poles and $\Mcal(t)$ is automatically a smooth function of $t$. We can see this in the right plot of Fig.~\ref{fig:sChanResiduals}, where $\Mcal(s)$ is a smooth, well-converged function for negative values of $s$.

\subsubsection*{\texorpdfstring{$u$}{u}-flavor amplitude}

Since we have chosen to only reduce particles 2 and 3 in our LSZ formula, there is no manifest crossing symmetry $p_1 \lra -p_3$ in Eq.~\eqref{eq:amp_arbitrary_flavor_app}. The calculation of the $u$-flavor amplitude
\be
\fr{1}{N} \Mcal^{ijji}(s,t) \simeq \Mcal(u),
\ee
is therefore inherently different than that of the previous amplitudes. In particular, there is no contribution to the $JJ$ terms which is finite as $N \to \infty$. Instead, the full amplitude comes from the $J'$ term,
\be
\Mcal(u) \simeq \fr{1}{N} \<\ve{p}_4^i|J'^{\,jj}(0)|\ve{p}_1^i\>.
\ee

To see this explicitly, we need to briefly remind ourselves of the structure of the mass-squared matrix
\be
\mathbb{M}^2 = \mathbb{M}_{\textrm{(CFT)}}^2 + \de\mathbb{M}_{(m)}^2 + \de\mathbb{M}_{(\lambda)}^2.
\ee
The CFT and mass term contributions are particle-number-preserving, while the quartic interaction has a piece which preserves particle number and one which mixes states whose particle numbers differ by 2. We can thus instead divide the operator $\mathbb{M}^2$ into the following two pieces,
\be
\mathbb{M}^2 = \mathbb{M}_{(n\to n)}^2 + \de\mathbb{M}_{(n\to n\pm 2)}^2.
\ee

As shown in the previous subsection, the matrix elements of $\mathbb{M}_{(n\to n)}^2$ are $\order(1)$, while the matrix elements of $\de\mathbb{M}_{(n\to n\pm 2)}^2$ are $\order(1/\sqrt{N})$. At large $N$, we can therefore first numerically compute the energy eigenstates of $\mathbb{M}_{(n\to n)}^2$ via truncation and then treat $\de\mathbb{M}_{(n\to n\pm 2)}^2$ as a perturbative correction to those eigenstates.

To be concrete, let's consider the $1$- and $3$-particle sectors. Diagonalizing $\mathbb{M}_{(n\to n)}^2$, we find the resulting set of energy eigenstates:
\be
|\ve{p}^i\>_{\textrm{free}}, \quad |M_\alpha^2;\ve{q}\> \otimes |\ve{q}'^{\,i}\>_{\textrm{free}},
\ee
i.e., the one-particle eigenstate is unchanged from the free one and the three-particle eigenstates consist of two-particle eigenstates plus a free spectator. We can then use old-fashioned perturbation theory to compute the leading perturbative correction to the one-particle eigenstate due to $\de\mathbb{M}_{(n\to n\pm 2)}^2$,
\be
| \ve{p}^i \> \simeq | \ve{p}^i \>_{\textrm{free}} + \!\!\! \sum_{\alpha\in\{n=2\}} \! \int \! \dd\ve{q} \, \dd \ve{q}' \, \fr{{}_{\textrm{free}} \<\ve{q}'^{\,j}| \otimes \<M_\alpha^2; \ve{q}|\de\mathbb{M}_{(\lambda)}^2| \ve{p}^i \>_{\textrm{free}}}{m^2 - M_{(n=3)}^2} \, |M_\alpha^2;\ve{q}\> \otimes |\ve{q}'^{\,j}\>_{\textrm{free}},
\ee
where $M_{(n=3)}^2$ is the invariant mass of the full three-particle state,
\be
M_{(n=3)}^2 \equiv \Big(\tfrac{M_\alpha^2 + q_\perp^2}{q_-} + \tfrac{q_\perp^{\prime2} + m^2}{q'_-}\Big)(q_-+q'_-) - (q_\perp+q'_\perp)^2.
\ee

\begin{figure}
    \centering
    \includegraphics[width=0.75\linewidth]{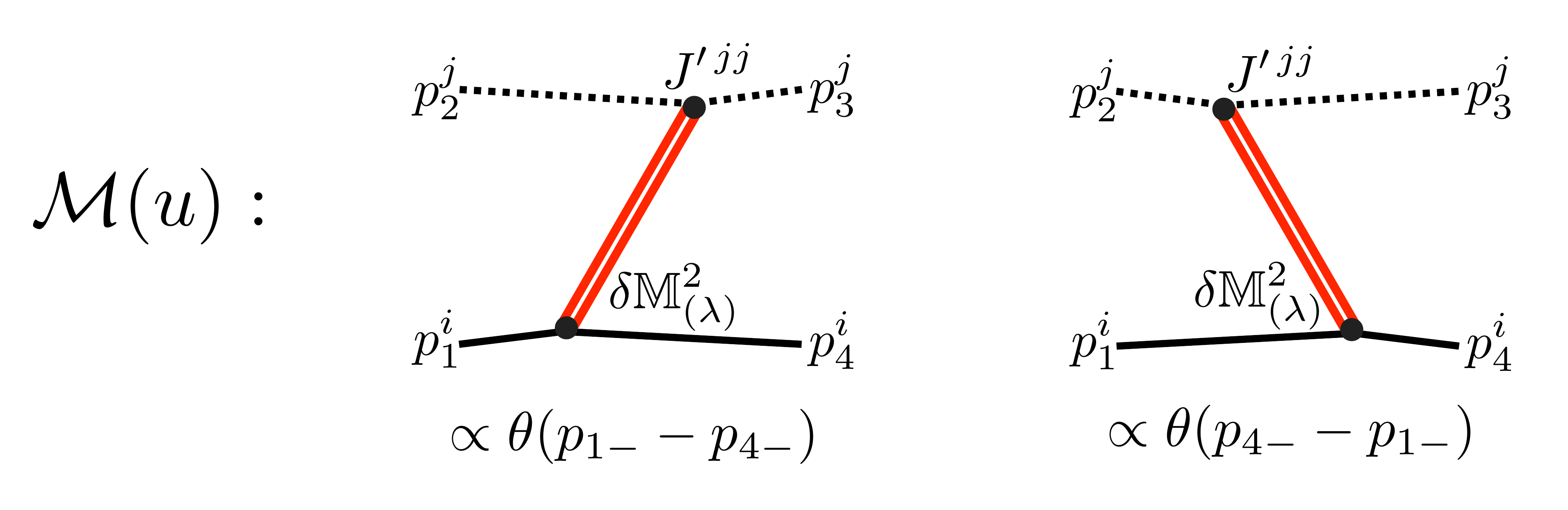}
    \caption{The $u$-flavor scattering amplitude receives contributions from the mixing of particle 1 (left) and particle 4 (right) with three-particle states containing two-particle flavor-singlet energy eigenstates and a single spectator.}
    \label{fig:uchannel_amp_detailed}
\end{figure}

Inserting this expansion for both external states in the $J'$ term, we obtain
\be\label{eq:UChannelJprime}
\ba
    \< \onmom{4}{i} | J'^{\,jj} | \onmom{1}{i} \> &\simeq \!\!\!\! \sum_{\alpha\in\{n=2\}} \! \int \! \dd\ve{q} \, \dd \ve{q}' \Bigg[ \frac{\big({}_{\textrm{free}}\< \onmom{4}{i} | J'^{\,jj} | M_\alpha^2 ; \ve{q} \> \! \otimes \! |\ve{q}^{\prime k}\>_{\textrm{free}} \big) \big({}_{\textrm{free}} \<\ve{q}^{\prime k}| \! \otimes \! \< M_\alpha^2 ; \ve{q} | \de \mathbb{M}_{(\lambda)}^2 | \onmom{1}{i} \>_{\textrm{free}}\big)}{m^2 - M_{(n=3)}^2}  \\
    & \hspace{2.5cm} + \frac{\big({}_{\textrm{free}}\< \onmom{4}{i} | \de \mathbb{M}_{(\lambda)}^2 | M_\alpha^2 ; \ve{q} \> \! \otimes \! |\ve{q}^{\prime k}\>_{\textrm{free}} \big) \big({}_{\textrm{free}} \<\ve{q}^{\prime k}| \! \otimes \! \< M_\alpha^2 ; \ve{q} | J'^{\,jj} | \onmom{1}{i} \>_{\textrm{free}}\big)}{m^2 - M_{(n=3)}^2} \Bigg] \\
    & \hspace{2.5cm} + {}_{\textrm{free}}\< \onmom{4}{i} | J'^{\,jj} | \onmom{1}{i} \>_{\textrm{free}}.
\ea
\ee
In addition to the same $\order(\lambda)$ term as the other flavor amplitudes, we have two contributions due to the mixing of particles 1 and 4 with three-particle states, shown schematically in Fig.~\ref{fig:uchannel_amp_detailed}.

Let's go through the first term, coming from the leading correction to particle 1, in detail. First we need to evaluate the $J'$ matrix element
\be
{}_{\textrm{free}}\< \onmom{4}{i} | J'^{\,jj} | M_\alpha^2 ; \ve{q} \> \otimes |\ve{q}^{\prime k}\>_{\textrm{free}} \simeq - \lambda \, \de^{ik} \de(\ve{p}_4 - \ve{q}') \<\vec{\phi}^{\,2}(0)| M_\alpha^2 ; \ve{q} \>.
\ee
Combining this with the $\de\mathbb{M}^2$ matrix element we computed in Eq.~\eqref{eq:MatrixElement1to3}, we obtain the $u$-flavor amplitude contribution:
\be
    \Mcal(u) \supset -\fr{\lambda^2}{N} \left( \fr{p_{1-}}{p_{1-} - p_{4-}} \right) \sum_{\alpha\in\{n=2\}} \frac{\<\vec{\phi}^{\,2}(0)| M_\alpha^2 ; \ve{p}_1 - \ve{p}_4 \> \<M_\alpha^2;\ve{p}_1 - \ve{p}_4|\vec{\phi}^{\,2}(0)\>}{m^2 - M_{(n=3)}^2}. 
\ee
We can simplify this expression by rewriting $M_{(n=3)}^2$ in the more useful form,
\be
M_{(n=3)}^2 = \Big(\tfrac{M_\alpha^2 + (p_{1\perp} - p_{4\perp})^2}{p_{1-} - p_{4-}} + \tfrac{p_{4\perp}^2 + m^2}{p_{4-}}\Big)p_{1-} - p_{1\perp}^2 = \left( \fr{p_{1-}}{p_{1-} - p_{4-}} \right) (M_\alpha^2 - u) + m^2,
\ee
such that we obtain
\be
    \Mcal(u) \supset \fr{\lambda^2}{N} \sum_{\alpha\in\{n=2\}} \frac{\<\vec{\phi}^{\,2}(0)| M_\alpha^2 ; \ve{p}_1 - \ve{p}_4 \> \<M_\alpha^2;\ve{p}_1 - \ve{p}_4|\vec{\phi}^{\,2}(0)\>}{M_\alpha^2 - u}. 
\ee
Analogously to the $JJ$ contributions for the previous amplitudes, this term is proportional to $\theta(p_{1-} - p_{4-})$, as indicated in Fig.~\ref{fig:uchannel_amp_detailed}.

We can repeat these manipulations for the second term in Eq.~\eqref{eq:UChannelJprime}, coming from the leading correction to particle 4, which gives an equivalent term proportional to $\theta(p_{4-} - p_{1-})$. Combining these various contributions together, we have the final expression
\be\label{eq:UChannel_Full}
    \Mcal(u) \simeq \fr{\lambda^2}{N} \sum_{\alpha\in\{n=2\}} \frac{\big\< \vec{\phi}^{\,2}(0) \big| M_\alpha^2 ; |\ve{p}_1 - \ve{p}_4| \, \big\> \big\< M_\alpha^2 ; |\ve{p}_1 - \ve{p}_4| \, \big| \vec{\phi}^{\,2}(0) \big\>}{M_\alpha^2 - u} - 2\lambda.
\ee
This expression is equivalent to the $s$- and $t$-flavor amplitudes computed above, with the crucial exception that it manifestly has no imaginary part. This is because $u$ is \emph{required} to take a physical value, unlike $s$ and $t$, due to the fact that particles 1 and 4 correspond to physical external states. Eq.~\eqref{eq:UChannel_Full} is therefore only defined for $u \leq 0$, such that we never encounter explicit poles and $\Mcal(u)$ is automatically a smooth function of $u$.

\section{Canonical derivation of central formula}
\label{app:LSZ_alt_derivation}
In the main text we derived the central formula Eq.~\eqref{eq:SDforLSZcorrelator} from a path integral perspective using Schwinger-Dyson relations. This formula can also be derived straightforwardly from a canonical perspective, which we detail here. Our starting point is the LSZ reduced amplitude, where under the Fourier transform the integrand contains
\begin{equation}
    D_3 D_2 \braket{p_4|T\{\phi(x_3)\phi(x_2)\}|p_1},
\end{equation}
and where we have introduced the notation
\[
D_i \equiv \Box_{x_i} +m^2, ~ \Box = 2\partial_+\partial_- - \partial_{\perp}^2.
\]
In the above we have displayed \(\Box\) in lightcone coordinates; throughout this appendix we choose to carry out computations in lightcone quantization (it is algebraically simpler since \(\Box\) is linear in the time derivative \(\partial_+\)), although the results can equally well be derived in other quantization schemes such as the standard equal-time quantization. In lightcone quantization the canonical momentum is given by
\[
\pi(x) = \frac{\partial \mathcal{L}(x)}{\partial\, (\partial_+\phi)} = \partial_- \phi(x),
\]
together with equal-time canonical commutation relations (CCR)
\begin{equation}
    [\phi(x^+,\mathbf{x}),\pi(x^+,\mathbf{y})] = \frac{i}{2}\delta^{(d-1)}(\mathbf{x}-\mathbf{y}),
\end{equation}
where we recall lightcone quantization has a funny factor of \(\frac{1}{2}\) in the CCR, i.e.\ \([q,p] = \frac{i}{2}\) in lightcone as opposed to \([q,p] = i\).

Returning to the correlation function entering the LSZ formula, we focus on 
\begin{equation}
    D_xD_y\, T\{\phi(x)\phi(y)\}, \nonumber
\end{equation}
and allow the differential operators to act on the time-ordered product
\begin{equation*}
T\{\phi(x)\phi(x)\} = \phi(x)\phi(y) \Theta(x^+-y^+) + \phi(y)\phi(x) \Theta(y^+-x^+).
\end{equation*}
Intuitively, when \(D_i\) hits a field we can use the equation of motion \(D_i\phi(x_i) = J(x_i)\), while contact terms can arise from the time derivative hitting theta functions in the time-ordering. We assume that $\phi$ and the source $J$ commute at equal times. Consider first
\begin{align}
    D_x T\{\phi(x)\phi(y)\} &= T\{D_x\phi(x) \phi(y)\} + 2 \delta(x^+-y^+) [\partial_-\phi(x),\phi(y)] \nonumber \\
    &= T\{J(x) \phi(y)\} - i \delta^{(d)}(x-y),
\end{align}
where in going to second equality we have used the EOM \(D_x\phi(x) = J(x)\), while the delta function in time allows us to use the CCR. Now acting on the above with \(D_y\) yields
\begin{align}
    D_yD_x T\{\phi(x)\phi(y)\} &= T\{J(x) D_y\phi(y)\} - 2 \delta(x^+-y^+)[J(x),\partial_-\phi(y)] -i D_y \delta^{(d)}(x-y) \nonumber \\
    &= T\{J(x)J(y)\} -i \frac{\delta J(x)}{\delta \phi(y)} -i D_y \delta^{(d)}(x-y). \label{eq:DDTphiphi}
\end{align}
In the second equality we have used the ``field representation" of the canonical momentum (the field generalization of \([q,p] = i\) implying the coordinate representation \(p = -i \frac{\partial}{\partial q}\)):
\[
2 \delta(x^+-y^+) \, [\pi(x),\mathcal{O}(y)] = -i \frac{\delta \mathcal{O}(y)}{\delta \phi(x)}.
\]
We see that Eq.~\eqref{eq:DDTphiphi} gives another derivation of Eq.~\eqref{eq:SDforLSZcorrelator} (the last term in Eq.~\eqref{eq:DDTphiphi} is a disconnected contribution which vanishes on-shell; it of course arises in the Schwinger-Dyson derivation, but was omitted in the main text for simplicity).

For completeness, let us a make a few other comments. LSZ is generally formulated in terms of time-ordered correlation functions; however, up to disconnected pieces, the time ordering can be replaced by retarded commutators (in fact, this is how LSZ was originally formulated~\cite{Lehmann:1954rq,Lehmann:1957zz}). Retarded commutators are occasionally useful for clarifying aspects of causality. Specifically, for our purposes, the scattering amplitude obtained from \(\braket{p_4|T\{\phi \phi\}|p_1}\) is equivalent to that obtained by replacing \(T\{\phi(x)\phi(y)\} \to R\{\phi(x)\phi(y)\}\), where the retarded commutator is given by
\begin{equation}
    R\{\phi(x)\phi(y)\} \equiv [\phi(x),\phi(y)]\Theta(x^+ - y^+) . \nonumber
\end{equation}
Acting with \(D_x\) and \(D_y\) gives the same formulas as above, with the replacement of time-ordering by the retarded commutator, \(T\{\cdots\} \to R\{\cdots\}\). In particular,
\begin{equation}
    D_yD_x R\{\phi(x)\phi(y)\} = R\{J(x)J(y)\} -i \frac{\delta J(x)}{\delta \phi(y)} -i D_y \delta^{(d)}(x-y) . \nonumber
\end{equation}
Repeating the steps that led to Eq.~\eqref{eq:GeneralAmp}, but this time with the retarded commutator, will again yield Eq.~\eqref{eq:GeneralAmp}, albeit with a single, minor change: the \(i\epsilon\) in the \(t\)-kinematics term flips sign, i.e.\ the \(1/(M_\alpha^2 - t - i\epsilon)\) piece is replaced by \(1/(M_\alpha^2 - t + i\epsilon)\).

\bibliography{bibliography}

\end{document}